\documentclass[twocolumn]{aastex62}
\pdfoutput=1
\usepackage{wasysym}
\usepackage{graphicx}
\usepackage{lineno}
\usepackage{xcolor}
\accepted{into ApJ}

\shorttitle{SUPERNOVA ORIGINS OF STARDUST}
\shortauthors{Schulte et al.}

\begin{document}

\title{THREE-DIMENSIONAL SUPERNOVA MODELS PROVIDE NEW INSIGHTS INTO THE ORIGINS OF STARDUST}

\correspondingauthor{Maitrayee Bose}
\email{Maitrayee.Bose@asu.edu}

\author{Jack Schulte}
\affil{Arizona State University, Department of Physics, 550 E Tyler Mall, PSF Rm \#470, Tempe, AZ 85287, USA}

\author{Maitrayee Bose}
\affiliation{Arizona State University, School of Earth and Space Exploration, 550 E Tyler Mall, PSF Rm \#686, Tempe, AZ 85287, USA}
\nocollaboration

\author{Patrick A. Young}
\affiliation{Arizona State University, School of Earth and Space Exploration, 550 E Tyler Mall, PSF Rm \#686, Tempe, AZ 85287, USA}
\nocollaboration

\author{Gregory S. Vance}
\affiliation{Arizona State University, School of Earth and Space Exploration, 550 E Tyler Mall, PSF Rm \#686, Tempe, AZ 85287, USA}
\nocollaboration

\begin{abstract}

We present the isotope yields of two post-explosion, three-dimensional 15 $M_\odot$ core-collapse supernova models, 15S and 15A, and compare them to the carbon, nitrogen, silicon, aluminum, sulfur, calcium, titanium, iron, and nickel isotopic compositions of SiC stardust. We find that these core-collapse supernova models predict similar carbon and nitrogen compositions to SiC X grains and grains with $^{12}$C/$^{13}$C $<$ 20 and $^{14}$N/$^{15}$N $<$ 60, which we will hereafter refer to as SiC 'D' grains. Material from the interior of a 15 $M_\odot$ explosion reaches high enough temperatures shortly after core collapse to produce the large enrichments of $^{13}$C and $^{15}$N necessary to replicate the compositions of SiC D grains. The innermost ejecta in a core-collapse supernova is operating in the neutrino-driven regime and undergoes fast proton capture after being heated by the supernova shockwave. Both 3-D models predict 0.3 $<$ $^{26}$Al/$^{27}$Al $<$ 1.5, comparable to the ratios seen in SiC X, C, and D grains. Models 15S and 15A, in general, predict very large anomalies in calcium isotopes but do compare qualitatively with the SiC X grain measurements that show $^{44}$Ca and $^{43}$Ca excesses. The titanium isotopic compositions of SiC X grains are well reproduced. The models predict $^{57}$Fe excesses and depletions that are observed in SiC X grains, and in addition predict accurately the $^{60}$Ni/$^{58}$Ni, $^{61}$Ni/$^{58}$Ni, and $^{62}$Ni/$^{58}$Ni ratios  in SiC X grains, as a result of fast neutron captures initiated by the propagation of the supernova shockwave. Finally, symmetry  has a noticeable effect on the production of silicon,  sulfur, and iron isotopes in the SN ejecta. 

% Removed these sentences to make the word count 253: Enrichments of $^{29,30}$Si seen in SiC C grains are the result of the propagation of the supernova shock through material with a moderate neutron excess. The symmetric model, 15S, overall does a better job at explaining the $^{33}$S and $^{34}$S excesses in anomalous SiC X, C and D grains. 

\end{abstract}


\section{Introduction} \label{sec:intro}

Supernovae (SNe) are some of the most energetic events in the universe. These events are capable of generating $10^{51}$ ergs, roughly equivalent to the total energy produced by the Sun throughout its 10 billion year lifetime, in approximately 100 seconds \citep{1995ApJS..101..181W}. The enormous heat generated by these stellar explosions is capable of producing a significant fraction of the elements heavier than carbon \citep{1957RvMP...29..547B}. As a result, much of the heavy elements that formed the terrestrial planets originated in the SNe that seeded the early solar system. While there are multiple ways a star can go supernova, this paper describes the more common variety, core-collapse supernovae (CCSNe). See \cite{2005NatPh...1..147W} for a review on this subject. Briefly, in a core-collapse supernova (CCSN), a star with a mass of greater than 8 times the mass of the Sun, after having already formed O, Mg, Si, S, Ar, and Ca in neon and oxygen burning, fuses silicon to form an iron core. Because iron has the largest nuclear binding energy per nucleon, fusion halts once an iron core has been produced. Neutrino cooling and other processes cause central pressures that keep a star from collapsing under its own gravity to rapidly decline. As a result, the core collapses either to a neutron star or black hole. This release of gravitational potential energy couples to the stellar material in a process that depends upon the compact object and progenitor characteristics, generating a shockwave that ejects the remaining stellar material and drives a significant amount of explosive nucleosynthesis. This ejecta contains substantial amounts of intermediate mass and Fe-peak isotopes with various neutron excesses and light r-process (rapid neutron capture) nuclei, signatures that are consistent with nucleosynthesis and mixing processes in CCSNe. The isotopic compositions of different parts of the ejecta depend on the location of the material within the progenitor star and its thermodynamic history during the explosion. The isotopic signatures of the dust grains produced in SN winds act as cosmic fingerprints that can be used to trace their nucleosynthetic histories and understand grain formation in massive stellar outflows.

The protosolar nebula that formed our solar system is derived from multiple generations of stellar ejecta. While most of this primordial material was destroyed or drastically altered in the formation of our solar system, a small amount of unaltered material remained. In 1987, the nanometer- to micron-sized dust particles were first isolated in meteorites \citep{1987Natur.330..728B}. These "presolar" (formed before the sun) dust grains, or stardust, exhibit large anomalies in their major and trace elements when compared to solar isotopic ratios \citep{2014mcp..book..181Z,2016ARA&A..54...53N,2016GeocJ..50....3F} and exhibit a range in chemical compositions. Mainstream SiC grains, Y, Z, and AB grains likely condensed in low-mass asymptotic giant branch and J-type carbon stars \citep[e.g.,][]{1991LPI....22..221C,1994ApJ...430..870H,1997ApJ...487L.101H,2000M&PS...35..997A,2003GeCoA..67.4961N,2005NuPhA.758..348N,2001NuPhA.688..102Z,2006ApJ...650..350Z}, while SiC X and C grains have been argued to condense in SNe \citep[e.g.,][]{1992ApJ...394L..43A,1993GeCoA..57.2869A,1996Sci...272.1314H,2000M&PS...35.1157H,2009GeCAS..73R.548H,2010ApJ...719.1370H,2012ApJ...745L..26H,2018GeCoA.221..182H,2002ApJ...575..257L,2003GeCoA..67.4961N,2010LPI....41.1359Z}. The origins, namely stellar site, associated nucleosynthesis, specific mechanisms and internal structures of the stars, are dictated by the isotopic signatures, as acquired by secondary ion mass spectrometry (SIMS) and more recently resonance ionization mass spectrometry (RIMS) instruments \citep[e.g.,][]{2005ApJ...631L..89N,2008ApJ...689..622M,2012ApJ...745L..26H,2015ApJ...799..156X,2016ApJ...820..140L,2003GeCoA..67.3215S,2018GeCoA.221..109S,2011PNAS..10819135Z}. 

The work presented here primarily focuses on a category of presolar grains that has large enrichments of $^{13}$C and $^{15}$N (hereafter called SiC D grains), characterized by $^{12}$C/$^{13}$C $<$ 20 and $^{14}$N/$^{15}$N $<$ 60, which have been connected to multiple origins \citep{2005ApJ...631L..89N,2016ApJ...820..140L,2019ApJ...873...14B,2001ApJ...551.1065A}. We also investigate the origins of SiC X grains, the largest population of SiC grains with possible CCSN links.  SiC X grains have large $^{15}$N excesses relative to solar material as a result of helium burning, accompanied by $^{12}$C/$^{13}$C ratios between 20 and 10,000 \citep{1991Natur.349...51Z,1992ApJ...394L..43A} as well as $^{28}$Si enrichments as a result of explosive oxygen burning. SiC C grains, a smaller population of grains that have similar carbon and nitrogen isotopic compositions to SiC D and SiC X grains but with large $^{28}$Si depletions are also a subject of this paper.

The connections between X grains and CCSNe were initially established through the use of one-dimensional (1-D) SN models \citep{2002ApJ...576..323R,1995ApJS..101..181W}. These models, while successful in explaining several isotopic trends seen in X grains, require an ad-hoc mixing of the zones in a pre-SN star, a circumstance which has little evidence of occurring, according to multi-dimensional SN models and observations of spatially resolved SN remnants \citep[e.g.,][]{2016ARNPS..66..341J}. Supernova explosions can produce large-scale mixing of chemically diverse zones \citep{2000ApJ...528L.109H}. Microscopic chemical mixing in the explosion is limited primarily to turbulent Kelvin-Helmholtz instabilities at the interface of structures, e.g., Rayleigh-Taylor fingers moving at different velocities. As the Kelvin-Helmholtz instability requires a small Richardson number (effectively a small entropy contrast), any unstable regions are very thin and only mix immediately adjacent regions \citep[i.e.][]{2003A&A...408..621K,2010ApJ...723..353J,2010MNRAS.407.1933J}. The total mass involved in microscopic mixing is very small. Additionally, it is physically impossible to do microscopic mixing between zones, e.g., one in the interior, a few in the middle, and others close to the hydrogen envelope. Finally, it is impossible to completely incorporate the hydrodynamic physics of supernova explosions in 1-D. The 3-D CCSN models presented here take into account asymmetries, turbulence, convection, and instabilities, allowing us to avoid such problems \citep{2009ApJ...699..938Y,2012ApJ...755..160E}. Thus, this paper, for the first time, presents realistic three-dimensional (3-D) CCSN models and their comparisons to SiC stardust compositions.

We show that 15 $M_\odot$ SNe are capable of producing material across the entire range of $^{12}$C/$^{13}$C ratios seen in presolar SiC grains, while simultaneously matching compositions of several other isotopic systems observed in most SN grains. Particularly, these models are capable of explaining $^{13}$C and $^{15}$N excesses, as well as $^{26}$Al/$^{27}$Al ratios measured in SiC D grains. We also investigate the differences between a spherically symmetric CCSN model and one that is velocity-asymmetric, an approach which cannot be taken using 1-D models, and find that symmetry has a noticeable effect on the silicon, sulfur, and iron isotopic composition of the ejecta. Using 3-D simulations allows for a better approximation of the complex processes in supernovae and provides more accurate nucleosynthetic yields so that better constraints may be made as to the origins of SiC stardust. This work therefore aims to prove the usefulness of 3-D SN models by using more physically accurate simulations to better constrain the origins of several different types of stardust grains and, as a result, provide insight into the energetic events that may have seeded the solar nebula and provided key elements to the terrestrial planets.

\section{Methods} \label{sec:methods}

The two 3-D simulations in this work used progenitors from the 1-D stellar evolution code TYCHO \citep{2005ApJ...618..908Y} to bring the star to core collapse, where the collapse was followed with a 1-D Lagrangian code \citep{1994ApJ...435..339H}. After core bounce, the SN shock propagates outward rapidly at first but strong neutrino cooling and photodissociation of the Fe peak nuclei causes the stalling of the shock front. The stalled shock waves was revived by an energy injection following the method laid out by \cite{2018ApJ...856...63F} where the energy is carried out by neutrinos and is deposited in the heating region.  Following shock revival the  star was then mapped into the 3-D smoothed particle hydrodynamics (SPH) code SNSPH \citep{2006ApJ...643..292F} \citep{2009ApJ...699..938Y}. The simulations are then post-processed using the Burnf code \citep{2007ApJ...664.1033Y} to obtain more accurate isotope yields. The 3D model calculation uses gray neutrino transport. Neutrino chemistry is included in the nucleosynthesis based on the fluxes from the 3D calculation. Test nucleosynthesis calculations on end member cases of lepton fraction plausible for these regions of the progenitor after neutrino processing confirm that in these conditions the gray approximation introduces smaller effects than the range of thermodynamic conditions encountered in the calculation. This range is in turn less than the uncertainties inherent in supernova calculations in general (See \cite{2020ApJ...895...82V} for a detailed discussion of the simulations).

%Stellar evolution calculations require tremendous amounts of CPU time and therefore cannot be done in 3-D. It is important, however, to follow the post-core-collapse explosion in 3-D in order to analyze the structure post-explosion and to compare with imaging observations \citep[e.g.,][]{2020ApJ...895...82V}. For these reasons, the simulation was done first in 1-D when computational limitations necessitated as such (stellar evolution until core-collapse), and then mapped into 3-D for the remaining evolution past core-collapse. 

In 3-D simulations and observed remnants, supernova ejecta are seen to form overdense, bounded clumps. These clumps largely maintain the composition of material from that location in the progenitor star, modified by explosive nucleosynthesis, without microscopic mixing from different regions. These clumps persist to large distances ($>$ parsec scales) in the supernova remnant. Calculations for different models, including the one referred to as 15S below, into the early remnant phase show that clumps maintain their integrity out to late times. Interactions of explosion-generated structures with structures arising from interations between stellar wind and CSM can, in fact, consolidate structures \citep{2013arXiv1305.4137E} In this work we will refer to clumps as the physical structure that is the location of dust formation and is consequently involved in delivery of material to the interstellar medium. As clumps in the simulation are represented by a number of  SPH particles with similar origins and thermodynamic histories, we will typically refer to clumps as a more physical interpretation of the ejecta, while SPH particles are the computational unit. Each SPH particle has an average mass of 3.2 $M_\oplus$.

Two simulations that use 15 $M_\odot$ progenitor stars were performed. One of these simulations, which we will call 15S (referred to in \cite{2019LPI....50.1746S,2020LPI....51.1268S} as 50Am), is a spherically symmetric explosion. The other, which we will call 15A (previously referred to as jet3b), has 2:1 velocity asymmetry introduced between the poles and the equator of the explosion. This asymmetry was introduced by increasing the velocities of the clumps within 30$^{\circ}$ of the z-axis by a factor of 6 and decreasing the velocities of the remaining clumps by a factor of 1.2 to conserve energy during the explosion \citep{2009ApJ...699..938Y}.

For comparison and completeness, we also use a 1-D post-explosion CCSN model from \cite{2002ApJ...576..323R}. This model was calculated using the presupernova implicit hydrodynamics package KEPLER from \cite{1978ApJ...225.1021W} accompanied by improved stellar physics (primarily the inclusion of mass loss) and two adaptive nuclear reaction networks. The supernova explosion is simulated by a piston model that moves inward to a radius of 500 km before being rebounded to a radius of 10,000 km. An abundance cutoff of $10^{-24}$ is used to capture isotope yields. Greater detail is discussed in \cite{2002ApJ...576..323R}. The two data sets used here, labeled 1D15C and 1D15Y in this work, are the zonal compositions and zonal yields from the model labeled S15 (internal name s15a28c) in \cite{2002ApJ...576..323R}. Each data set represents isotopic information from a 15 $M_\odot$ explosion approximately seven hours after core bounce. 1D15C represents the isotopic compositions of all 'zones' at a given mass coordinate (or radius) from the center of the explosion. 1D15Y takes into account the contributions of radioactive progenitor isotopes (such as $^{26}$Al) when calculating the abundances of isotopes that are decay products.

The 3-D model output is in the form of text files which consist of the X-Y-Z position of each clump approximately 43 hours after the explosion (after which no more burning is expected to occur), followed by the total mass, smoothing length, density, and mass fraction of all relevant isotopes (abundance cutoff at 10$^{-10}$). Additional output files include the temperature and density histories of selected clumps. These data files are then processed using MATLAB, where relevant isotope ratios are computed, distance scales are converted, and the output is analyzed. Additionally, we used the SPH visualization tool SPLASH \citep{2007PASA...24..159P} to complement MATLAB in generating planar abundance maps and density maps of the explosions. Preliminary results were reported in \cite{2019LPI....50.1746S} and \cite{2020LPI....51.1268S}.

The stardust isotopic data used in this work were obtained from the presolar grain database, which is maintained by the Washington University in St. Louis. This database has been updated significantly \citep{2020LPI....51.2140S} since its inception in 2009 \citep{2009LPI....40.1198H}, with respect to isotopic measurements of SiC grains.

\section{Results and Discussion} \label{sec:results}
\subsection{Temperature and Density History of CCSN Material}

We first investigated the temperature and density of the material in the interior of the SN as a function of time. These temperatures and densities are correlated and decay according to a power law, as a function of time elapsed after core collapse. Shortly after core collapse, material close to the center of both 3-D models reaches temperatures greater than $3 \times 10^9$ K (Fig. 1a). Ejecta in the interior of the explosion has a density of $> 3 \times 10^5$ g cm$^{-3}$ initially (Fig. 1b). 

In the first 10 seconds after core bounce, the innermost ejecta remains at a temperature greater than $3 \times 10^8$ K (Fig. 1a), under which conditions rapid proton capture is possible \citep{1998PhR...294..167S}, which produces an excess of the key isotope $^{13}$C (Fig. 2), leading to a principal difference between SiC D and SiC X grains. Material that later forms SiC X grains reaches an average temperature of $3 \times 10^8$ K for only a brief period at most, leading to limited $^{13}$C production. The increased flux of neutrinos that follows core bounce also enables the neutrino-induced production of $^{15}$N, $^{26}$Al, and $^{47,49}$Ti \citep{1990ApJ...356..272W} in the innermost ejecta, all isotopes which are relevant to the classification of presolar SiC grains.

43 hours after core collapse, 15S predicts that this material has traveled 1.1 AU from the center of the explosion and decreased in temperature and density to $4.7 \times 10^8$ K and $1.2 \times 10^3$ g cm$^{-3}$. The velocity-asymmetric explosion 15A predicts that this material of similar composition has traveled 1.4 AU from the center of the explosion, in the direction of the +Z pole, leading to the temperature decreasing to $7.4 \times 10^4$ K and the density decreasing to $3.6 \times 10^{-10}$ g cm$^{-3}$ in the same amount of time. The extremely high temperatures (~3-8 billion K) and pressures of the clumps being ejected from the heart of the SNe are conducive to the activation of the CNO cycle, which is capable of producing the stable nuclides of carbon and nitrogen \citep{1999JPhG...25R.133W}. The final isotope yields in the innermost ejecta, however, are dictated by the interplay of the neutrino reactions, and alpha and neutron capture sequences \citep{2005ApJ...623..325P,2006AIPC..847..333F,2011ApJ...729...46W,2016ApJ...833..124M}.

\subsection{Comparisons to SiC Isotopes}
\subsubsection{Carbon and Nitrogen}

Historically, the isotopes $^{12}$C, $^{13}$C, $^{14}$N, and $^{15}$N were studied in the greatest detail because they, along with the isotopes of silicon, are the best diagnostic tool for examining the origins of SiC stardust.

As is shown in Fig. 2, both the 15S and 15A simulations explain the carbon and nitrogen compositions of most SiC grains with $^{15}$N excesses ($^{14}$N/$^{15}$N $<$ 272) and a large range in carbon isotopic compositions (1 $<$ $^{12}$C/$^{13}$C $<$ $10^4$). All of the clumps within this range of $^{12}$C/$^{13}$C have $^{15}$N excesses consistent with SiC X, D, and C grain measurements and require no mixing between zones in the pre-SN star or with material from the interstellar medium or solar nebula, which has otherwise been a requirement of 1-D classical nova and SN models \citep{2002ApJ...576..323R} For instance, the classical ONe nova models used to explain SiC D grains require mixing models to simulate very large contributions (about $90$\%) from Solar matter, meaning that the minority of the material from which the grain formed is expected to come from a nova \citep{2001ApJ...551.1065A}. More recent CO nova models require much less contribution from Solar matter \citep{2019ApJ...873...14B} but mixing cannot be completely avoided.

Next, we were interested in determining the location in the CCSN explosion where $^{13}$C is being produced. To achieve this, the clumps generated by both models were split into two separate categories: first, the $^{13}$C enriched clumps, or those with $^{12}$C/$^{13}$C less than 20 (described as SiC D grains). The clumps that are more closely related to X grains, with $^{12}$C/$^{13}$C between 20 and 30,000, formed the remaining category. Our analysis shows that clumps with $^{13}$C and $^{15}$N excesses originate deep within the interior of the SN explosion (Fig. 3a, 3e, 3i), in both the asymmetric and symmetric models. With temperatures greater than $3 \times 10^8$ K and densities larger than 300 g cm$^{-3}$ until 10 seconds after the explosion, these clumps form with large $^{13}$C enrichments as a result of rapid proton capture, $^{12}$C($p$,$\gamma$)$^{13}$N($\beta^+$)$^{13}$C, being carried out near the center of the explosion. Photons emitted from the core dissociate the iron atoms into helium nuclei, prior to core bounce, resulting in a nucleon gas rich in alpha particles surrounding the proto-neutron star. Thus, corresponding enrichments in $^{15}$N are the results of these high temperatures and densities creating a free particle flux (alpha and proton) that enables the CNO cycle and therefore the reaction $^{14}$N$(\alpha,\gamma)^{18}$F$(p,\alpha)^{15}$O$(\beta^+)^{15}$N. 

All the clumps in the interior of each simulation that have $^{12}$C/$^{13}$C $<$ 20 also have $^{14}$N/$^{15}$N $<$ 60 (Fig. 3b, 3f, 3j). In general, for clumps interior to about 1.4 AU, the closer a clump is to the direct center of the explosion, the greater the magnitude of $^{15}$N enrichment is, possibly further enhanced by alpha capture on $^{14}$N nuclides. The interior of both the asymmetric and symmetric CCSN explosions consists almost entirely of material with $^{12}$C/$^{13}$C and $^{14}$N/$^{15}$N  $<$ 100. 

These novel 3-D CCSN simulations indicate that, through explosive proton and alpha capture, the same SN event can produce both SiC X and SiC D grains. It is possible for stardust with SiC X grain carbon and nitrogen compositions to form first in the outer shells of a 15 $M_\odot$ SN, while stardust grains with $^{13}$C and $^{15}$N enrichments, compositions similar to SiC D grains, are produced in the interior of the same explosion. This is a significant result when contrasted with 1-D SN simulations (Fig. 2) that predict $^{15}$N enrichments between $10^{-2}$ and 1 but only partially match the carbon and nitrogen compositions of SiC D grains. Our 3-D simulations, which use progenitor stars of the same mass as the 1-D model, predict up to 100 times as much $^{14}$N, matching the entire range in $^{14}$N/$^{15}$N ratios observed in SiC X, C and D grains.

\subsubsection{Aluminum}

Both 3-D CCSN simulations predict that 0.25\% of the simulated clumps will have 0.3 $<$ $^{26}$Al/$^{27}$Al $<$ 1.5 (Fig. 4). Symmetry in the 3-D models does not make a difference in the production of Al with the data from both simulations completely overlaying over the entire range of carbon isotope ratios. The ejecta composition agrees with the upper limit of the aluminum ratios observed in SiC X, D, and C stardust, and exhibits decreasing $^{26}$Al/$^{27}$Al ratios with decreasing $^{12}$C/$^{13}$C ratios. This trend qualitatively matches the compositions of the SiC D grains; i.e., aluminum ratios are smaller (0.3 to 0.6) in clumps with larger enrichments of $^{13}$C and increase towards $^{26}$Al/$^{27}$Al = 1.5 in clumps with carbon and nitrogen ratios which are comparable to SiC X grains (Fig. 4). 

Contrary to 3-D models, the 1-D CCSN simulation used for comparison from \cite{2002ApJ...576..323R} predicts much smaller $^{26}$Al/$^{27}$Al ratios (Fig. 4). These match the observations of SiC C and D grains with low $^{12}$C/$^{13}$C, but have $^{26}$Al/$^{27}$Al between $1 \times 10^{-4}$ and $6 \times 10^{-3}$ in ejecta with larger carbon isotope ratios, which is at least an order of magnitude smaller than what is expected for SiC X stardust.

The large implied abundances in the short-lived radionuclide aluminum-26 that are observed in SiC X, D, and C grains as well as in our CCSN models are the results of proton capture \citep{2014mcp..book..181Z,2000M&PS...35.1157H,2010ApJ...709.1157L,2015ApJ...799..156X}. $^{26}$Al is an rp-process isotope and the reaction leading to $^{26}$Al enrichments in SN explosions is $^{25}$Mg$(p,\gamma)^{26}$Al. This occurs at C and Ne burning conditions and at very high temperatures and densities where free particle fluxes are high, which are achieved near the center of the explosion (Fig. 3c, 3g, 3k). As a result, some of the material with the largest abundances of $^{26}$Al comes from near the center of a CCSN, in the same region as $^{13}$C and $^{15}$N are produced in large quantities. Therefore, the observed carbon, nitrogen and aluminum isotopic pattern in the grains suggests that high temperature ejecta in core-collapse supernovae can provide key source material for supernova SiC D grains.

\subsubsection{Silicon}

The 3-D simulations predict that two different populations of clumps will be ejected from a 15 $M_\odot$ CCSN. One set with enrichments of the isotope $^{28}$Si, which are only produced in the interior of the CCSN explosion, and another with depletions of $^{28}$Si, which are produced both in the interior of the CCSN as well as the exterior. Those with enrichments of $^{28}$Si have $\delta ^{30}$Si and $\delta ^{29}$Si as low as $-1000\permil$ each and therefore fail to precisely match the compositions of SiC X grains (Fig. 5). Those clumps with a depletion of $^{28}$Si match the Si compositions of SiC C grains in the range of $\delta ^{29}$Si = $600-2800 \permil$. The symmetric model 15S explains SiC C grains with larger enrichments of $^{29}$Si ($\delta ^{29}$Si = $1000 - 2800 \permil$) while the asymmetric model explains SiC C grains with smaller enrichments of the same isotope ($\delta ^{29}$Si = $600 - 1700 \permil$). Neither the 3-D nor 1-D models can predict the silicon isotope ratios of SiC D grains because they exhibit $^{30}$Si enrichments.

$^{28}$Si enrichments in the interior of a CCSN are the results of explosive oxygen burning \citep{2002RvMP...74.1015W,2013ApJ...779..123P,2018SciA....4.1054L}. This is a straightforward reaction in which two $^{16}$O nuclides fuse to form a $^{28}$Si nuclide and an $\alpha$ particle ($^{16}$O $+$ $^{16}$O $\rightarrow$ $^{28}$Si $+$ $\alpha$). Because this is a late-stage reaction that requires very high temperatures and densities, only clumps close to the center of the explosion will have $^{28}$Si excesses. Clumps with $^{29,30}$Si enrichments that are produced throughout the CCSN explosion are most likely caused by captures onto nuclei with a neutron excess. Thus, clumps with silicon compositions most similar to SiC X grains are produced in the interior of a CCSN explosion and experience explosive oxygen burning, which produces large amounts of $^{28}$Si, while clumps with Si compositions most similar to SiC C grains form throughout a CCSN explosion from neutron-rich isotopes that are not diluted by excess $^{28}$Si production. 

We argue that the SiC D grains formed in the innermost material in a CCSN explosion, and the carbon, nitrogen and aluminum isotopic compositions are a testament to that fact. No dilution of the original carbon and nitrogen isotopic signatures is necessary. However, to explain silicon isotopes, we may need to invoke a scenario where grain compositions changed as the grains nucleate and grew in the SN winds. The discrepancies between modeled Si yields and SiC D grains (Fig. 5) may be a result of the silicon in the SiC grains formed in CCSN ejecta being sputtered away in the ISM. SiC grains do experience preferential sputtering of silicon as they transit through the interstellar medium (ISM) and before being incorporated in the protosolar nebula \citep[e.g.,][]{1997A&A...321..293S}. Helium impacts from the ISM with velocities greater than 55 km/s (or as low as 25 km/s for heavier species) can release a significant amount of silicon from the 'skin' of a stardust grain into gaseous silicon, which forms much of the SiO observed in the ISM. However, while the outer skin of SiC grains is constantly recycled by the ISM, the bulk of the grain and its stoichiometry likely remain the same \citep{2003ApJ...594..312D}. Alternatively, SiC D grains show imprint of the dominant neutron capture process, depending on temperature and location of the seed nuclei, which leads to abundant $^{30}$Si production just before freeze-out. It seems, therefore, that the mismatch between the measured and modeled Si isotope signatures in the D grains continues to be an enigma.

\subsubsection{Sulfur}

Both CCSN simulations predict, as a result of oxygen and silicon burning \citep{1957RvMP...29..547B}, enrichments of the isotope $^{32}$S as large as $1000\permil$ in the exterior of the explosions (Fig. 6). Clumps enriched in the other two stable isotopes of sulfur, $^{33}$S and $^{34}$S, are only produced deep within the interior of the CCSN and, 43 hours after the explosion, have only been ejected at most 1.4 AU from the center of the explosion. The neutron-rich innermost ejecta from the CCSN is exposed to this late burning phase, causing these clumps to be enriched in $^{33}$S and $^{34}$S, while all of the clumps outside of 1.4 AU are still composed of mostly $^{32}$S. These broad observations are true in both the case of the asymmetric explosion and the symmetric explosion. There are few differences imposed by symmetry between the two 3-D models other than that the symmetric model seems to better predict the sulfur compositions of several anomalous SiC X grains.

The symmetric model 15S produces slightly more $^{33}$S, leading to $^{33}$S enrichments of up to $600\permil$ (Fig. 6). These predictions precisely match the compositions of several anomalous SiC X stardust grains that have $^{33}$S enrichments between 100 and 300$\permil$ \citep{2015ApJ...799..156X}. Most other SiC X, D, and C grains have sulfur compositions between that of the solar system and $600\permil$ $^{32}$S enrichments \citep{2012ApJ...745L..26H,2012LPI....43.2679O}, and the 3-D CCSN simulations explain all of these compositions with the exception of two SiC C grains, M7-C and M7-D, from \cite{2012ApJ...745L..26H}. Both of these grains show sulfur isotopic compositions that are predicted qualitatively by 15A and 15S models. They do not, however, explain these grain compositions quantitatively, which is odd considering that these grains with the largest $^{32}$S excesses also show clear evidence for $^{26}$Al and $^{44}$Ti decay. Their $^{26}$Al/$^{27}$Al ratios are within the range observed by the CCSN simulations. Since the $^{44}$Ca excess is clearly indicative of the CCSN origin in various types of presolar grains \citep{1996Sci...272.1314H,1996ApJ...462L..31N} and a SN spinel grain \citep{2010ApJ...717..107G}, additional models will attempt to reproduce these grain compositions with the largest $^{32}$S enrichments observed to date. 

It is important to note that, while the 3-D CCSN simulations do accurately predict the sulfur compositions of SiC D grains, only three such grains have been measured for sulfur, each with a different predominant isotope and with errors between 100 and 400$\permil$ \citep{2016ApJ...820..140L,2019ApJ...873...14B}. Terrestrial contribution to the sulfur isotopic compositions of SiC stardust grains is variable. The low sulfur concentrations in the SiC grains also produces large errors in the isotopic measurements. In spite of these experimental challenges, additional measurements of SiC D grains should be made to establish the sulfur isotopic compositions of these rare stardust grain types, especially since stardust preserves a record of sulfur molecule chemistry in the SN ejecta \citep{2012ApJ...745L..26H}, and sulfur isotopes are thought to be diagnostic of the stellar source \citep{2014PhLB..737..314P}.

\subsubsection{Calcium and Titanium}

There is little difference in the production of calcium or titanium isotopes between the symmetric and asymmetric simulations. Both 15S and 15A generally produce clumps with extremely large $^{42,43,44}$Ca excesses normalized to $^{40}$Ca, ranging from roughly solar values to several thousand permil. Here, we discuss the range of simulated values only in the range of anomalies noted in SiC stardust. No measurements have yet been done of calcium isotopes in SiC D grains, and as such, no implications can be made about the origins of SiC D grains. Calcium, like $^{32}$S, is a product of late silicon burning, the last major burning phase before explosion \citep{1957RvMP...29..547B}. In the CCSN models discussed here, a large fraction remains in the form of $^{32}$S.

Several SiC X grains and a SiC C grain have been measured for $^{44}$Ca \citep{1992ApJ...394L..43A,1996ApJ...462L..31N,2002ApJ...575..257L,2003GeCoA..67.4693B,2010ApJ...719.1370H,2012ApJ...745L..26H,2015ApJ...799..156X}. Enrichments of $^{44}$Ca, the stable product of the short-lived radionuclide $^{44}$Ti, are predicted by both models. A large section of these SiC X grains and the lone SiC C grain have enrichments of $\delta ^{44}$Ca/$^{40}$Ca values as large as 5270$\permil$ in SiC X grains and 1854$\permil$ in SiC C grains \citep{2003GeCoA..67.4693B}. Remaining grains show depletions in $^{44}$Ca, albeit with large errors \citep[e.g.,][]{Besmehn2001}. Both the CCSN simulations show clumps with $^{44}$Ca excesses greater than 1000$\permil$, as well as clumps that have $^{44}$Ca depletions of about -1000$\permil$. These same clumps also show $\delta ^{42}$Ca/$^{40}$Ca values greater than -1000$\permil$ but none of the stardust grains show $^{42}$Ca anomalies. Stardust grains with enrichments in $^{44}$Ca are likely to have condensed from material from the interior of the CCSN, where $^{44}$Ti is produced in large quantities (Fig. 3d, 3h, 3l). Grains with depletions of $^{44}$Ca are more likely to condense out of material from the exterior of the same explosion. Only three measurements have been done of the minor isotope $^{43}$Ca that show that SiC X grains typically have small excesses of this isotope (up to +500$\permil$). Neither of the 3-D CCSN models exactly match the grains' $^{43}$Ca/$^{40}Ca$ ratios but clumps with $\delta ^{43}$Ca values greater than 0$\permil$ do exist in both CCSN models.

Both simulations predict excesses of the neutron-rich titanium nuclides $^{49}$Ti and $^{50}$Ti greater than about 50$\permil$. These are accompanied by roughly solar $^{46,47}$Ti abundances when normalized to $^{48}$Ti. Titanium measurements of SiC X and D grains show that SiC X grains have enrichments in $^{50}$Ti up to 600$\permil$ \citep{1996ApJ...462L..31N} and both SiC X and D grains have enrichments in $^{49}$Ti between 0 and 1000$\permil$ \citep{2005ApJ...631L..89N,1996ApJ...462L..31N}. The titanium isotope anomalies in SiC D grains are small or not present so much of the comparison was done with the X grains. A minority of the clumps that form SiC X grains in the interior of the SN have large enrichments of $^{47,49}$Ti and therefore show evidence of rapid neutron capture, which can now explain the $^{47}$Ti excess observed in grain KJGM2-243-9 \citep{1996ApJ...462L..31N}. Both 15S and 15A match the SiC X grain compositions with very few exceptions: (1) $^{48}$Ti depletions in grains KJGM2-66-3 \citep{1996ApJ...462L..31N} and KJGN2-345-1 \citep{2003M&PSA..38.5118A} (2) $^{46}$Ti depletion in grain M9-137-11 \citep{2003GeCoA..67.4693B}. Although the current CCSN simulations do not explain the $^{46,47}$Ti characteristics, clumps with $\delta ^{50}$Ti/$^{48}$Ti of -300$\permil$ and $^{49}$Ti/$^{48}$Ti of -800$\permil$ are predicted in the simulations and thus can explain the $^{48,49,50}$Ti compositions of grain KJGM2-243-9  \citep{1996ApJ...462L..31N}.

\subsubsection{Iron and Nickel}

15S and 15A both predict large $^{57}$Fe enrichments ($>320\permil$; Fig. 7a). While the vast majority of the SPH clumps produced by the models have these enrichments, about 6\% of the clumps have depletions in $^{57}$Fe of about 920 to 980$\permil$ (Fig. 7a). These $^{57}$Fe depletions are expected to be the consequence of silicon burning, leading to the radioactive decay of $^{56}$Ni: $^{56}$Ni($\beta^+$)$^{56}$Co($\beta^+$)$^{56}$Fe. Both simulations also predict enrichments of the three neutron-rich isotopes of Ni: $\delta ^{60}$Ni/$^{58}$Ni $>$ 40\permil, $\delta ^{61}$Ni/$^{58}$Ni $>$ 220\permil, and $\delta ^{62}$Ni/$^{58}$Ni $>$ 75$\permil$ (Fig. 7b, 7c). These enrichments are most likely as a result of neutron capture associated with the propagation of the CCSN shockwave through neutron rich material in the inner regions of the progenitor.

No SiC D and C  grains have been measured for Fe-Ni isotope systematics but SiC X grains usually show enrichments in $^{57}$Fe between 0 and 1000 $\permil$ along with roughly solar $^{54}$Fe compositions \citep{2008ApJ...689..622M} (Fig. 7a). These grains have been explained with 1-D nucleosynthesis models by mixing material from the He/N zone with material from the He/C zone of the SN \citep{2008ApJ...689..622M}. Our 3-D CCSN models do not require any mixture between zones or with the protosolar nebula to explain the $^{57}$Fe enrichments in these grains. Additionally, four SiC X grains have been measured which contain depletions in $^{57}$Fe and cannot be explained by 1-D mixing models \citep{2008ApJ...689..622M} but can be qualitatively explained by our 3-D simulations. The clumps in both simulations show $\delta ^{57}$Fe/$^{56}$Fe values greater than 320 \permil. SiC X grains have also been consistently measured to have large excesses in $^{61}$Ni, and $^{62}$Ni with respect to solar ratios \citep{2008ApJ...689..622M}. These excesses are predicted by 15S and 15A, but the mostly solar $^{60}$Ni/$^{58}$Ni ratios ($\pm 100\permil$) which are predicted by the 1-D model do not show up in the 3-D simulations. The 3-D models do, however, match the large $^{60}$Ni, $^{61}$Ni, and $^{62}$Ni enrichments discovered in the grain M08-249-8 from \cite{2008ApJ...689..622M} (Figure 7b, 7c).

\subsection{Elemental Abundances of the Ejecta}

We calculated the abundances of key species, namely C, N, O, and Si, during SiC X and D grain condensation and computed C/O ratios of the ejected material. Note that the ejected material has an artificial abundance cutoff of 10$^{-10}$ to limit file sizes.

All clumps with $^{12}$C/$^{13}$C $<$ 30,000 have C/O $<$ 0.18. A fraction of grains with $^{12}$C/$^{13}$C $>$ 42,100 have C/O $>$ 31, but these grains have far too great $^{12}$C excesses to explain SiC X and D grain compositions. So in essence we are discussing SiC grain formation in an O-rich environment, which is viable. In the ionized ejecta of a supernova explosion, the atomic carbon vapor pressure is higher than expected under thermodynamic equilibrium at a temperature of ~1000 K, which can result in rapid condensation into carbon-rich dust grains \citep{1998LPI....29.1016C,1999Sci...283.1290C,2016LPI....47.2336M}. This occurs because the compositions of dust that forms in supernovae are dictated by kinetic chemistry rather than by equilibrium chemistry. Therefore, the requirement often imposed while using 1D supernova models is not a strict one, and condensation of carbon-rich dust grains will occur in oxygen-rich environments. Corroborating evidence was found via high resolution mapping  of SN 1987A by the Atacama Large Millimeter/Submillimeter Array (ALMA) that clearly show the production of abundant dust in the cold inner ejecta of the supernova remnant, and nearly all of the carbon has condensed into dust \citep{2014ApJ...782L...2I}. Therefore, condensation of carbon-rich species occur through an efficient pathway.

The model calculations presented in this paper have not been carried out to the point of grain condensation but we calculated the total yields of essential elements that would form SiC dust. About a quarter of the total carbon and 80\% of the total nitrogen in the explosion goes into the production of grains that match the carbon and nitrogen compositions of SiC X grains. Included in this material is 10\%  of the explosion's oxygen and 0.2\%  of the explosion's silicon. Consistent with our expectations, material that forms SiC D grains is much less abundant. When compared with the total elemental abundances in the supernova, material comparable to SiC D grains compose about 9 ppb of the explosion's carbon, 500 ppm nitrogen, 200 ppb oxygen, and 30 ppm silicon. Thus the abundance of SiC X grains ought to be much higher than SiC D grains. However, the abundances of X and D grains in the ejecta cannot be directly compared to those observed in the meteorite record because of dust destruction processes in the asteroid parent bodies \citep{1995GeCoA..59..115H}, during solar system formation and in the interstellar medium \citep[e.g.,][]{2004ASPC..309..347J}. The abundance of SiC X grains in Murchison is 0.063 ppm \citep{2014AIPC.1594..307A}. Based on the Si abundances for X (0.2 \%) and D grains (30 ppm) in the 15S CCSN models and assuming that all the Si is locked up in SiC grains during grain formation, we can infer that the SiC D grain abundance in meteorites is likely at least 2 orders of magnitude smaller than X grains.

SiC D grain production in the inner ejecta of the CCSN was deciphered, only as a result of using the state-of-the-art 3D models. The 3D models have allowed us to track the conditions for specific parts of the star. As a result, arbitrary, non-physical mixing between separate parts of the star are excluded. All of the physical conditions under which burning can occur, including complex thermodynamic trajectories resulting from 3D fluid motions, can be sampled. In some burning regimes a small entropy, temperature, or density change (Figure 1) can result in a significant change in the flux of protons, neutrons and/or alpha particles, and therefore isotopic yields. The high entropy bubbles experience proton capture within 10 s of core bounce. The high entropy bubbles have faster freezeout timescales, leading to the quenching of the nuclear reactions. Faster freeze-out timescales will result in higher light isotope abundances in general, which likely occurred for the freeze-out of SiC D grain precursor material. These bubbles are to some extent confined by higher density bubble walls, and their interaction with the complex circumstellar medium leads to the formation of complex structures \citep{2011PhDT........47E,2013arXiv1305.4137E}. The high entropy bubbles will, however, remain low density relative to other ejecta. As a result, SiC D grains should be statistically smaller in size than other supernova grains.

\section{Conclusions} \label{sec:conclusions}

We compare isotopic abundances from 3-D CCSN simulations to presolar grain compositions. The advantage of using 3-D simulations is that they preserve the unique thermodynamic histories of different parts of the ejecta, which can influence isotopic ratios, and do not impose nonphysical mixing. We have inferred that presolar SiC D with large $^{13}$C and $^{15}$N excesses, can condense out of material from the interior of a core-collapse SN as a result of proton and alpha capture in a high-temperature, high-density environment. SiC X grains form from material in the exterior and interior of the SN and show evidence of rapid neutron capture. The 3-D models shown in this work predict large abundances of the neutron-rich isotopes $^{29,30}$Si, $^{33,34}$S, $^{42,43,44}$Ca, $^{49,50}$Ti,$^{57}$Fe, and $^{60,61,62}$Ni as a result of nucleosynthesis happening after core collapse. Although SiC D grains are likely small and low in abundance, measurement of r-process nuclides in SiC D grains (currently unavailable) is capable of providing essential constraints on the CCSN engine.

While these models are the first to accurately reproduce the carbon and nitrogen compositions of SiC D grains and reproduce the trends observed in aluminum, sulfur, calcium, and titanium isotopes fairly well, they fall short of being able to explain the silicon isotope compositions of the same grains. While it is possible that admixture with large amounts of material in the stellar envelope or in the protosolar nebula will explain these discrepancies, further work will need to be done to understand such mixing processes during grain condensation and growth, and destruction processes in the interstellar medium. Thus, we show in this work how \textit{fresh} dust grains may condense from SN material, as the shock wave creates conditions optimal for the production of many of the isotopes present in stardust grains. 

In addition to finding that a rare fraction of stardust from the very interior of CCSN are available in meteorites, we investigated the differences between a spherically symmetric CCSN model and one that is velocity-asymmetric, an approach which cannot be taken using 1-D models. We infer that symmetry has a noticeable effect on the production of silicon, sulfur, and iron isotopes in the SN ejecta.

\section{Acknowledgements} \label{sec:acknowledgements}

Funding for the work was provided by startup funds to MB from Arizona State University. We would also like to thank the NASA/AZ Space grant program for providing partial funding to the undergraduate student, Jack Schulte. Support for this work was provided by the National Science Foundation award 1615575. We thank an anonymous referee for helpful comments, which helped to improve the original manuscript.

\begin{figure}[p]
  \centering
  \gridline{\fig{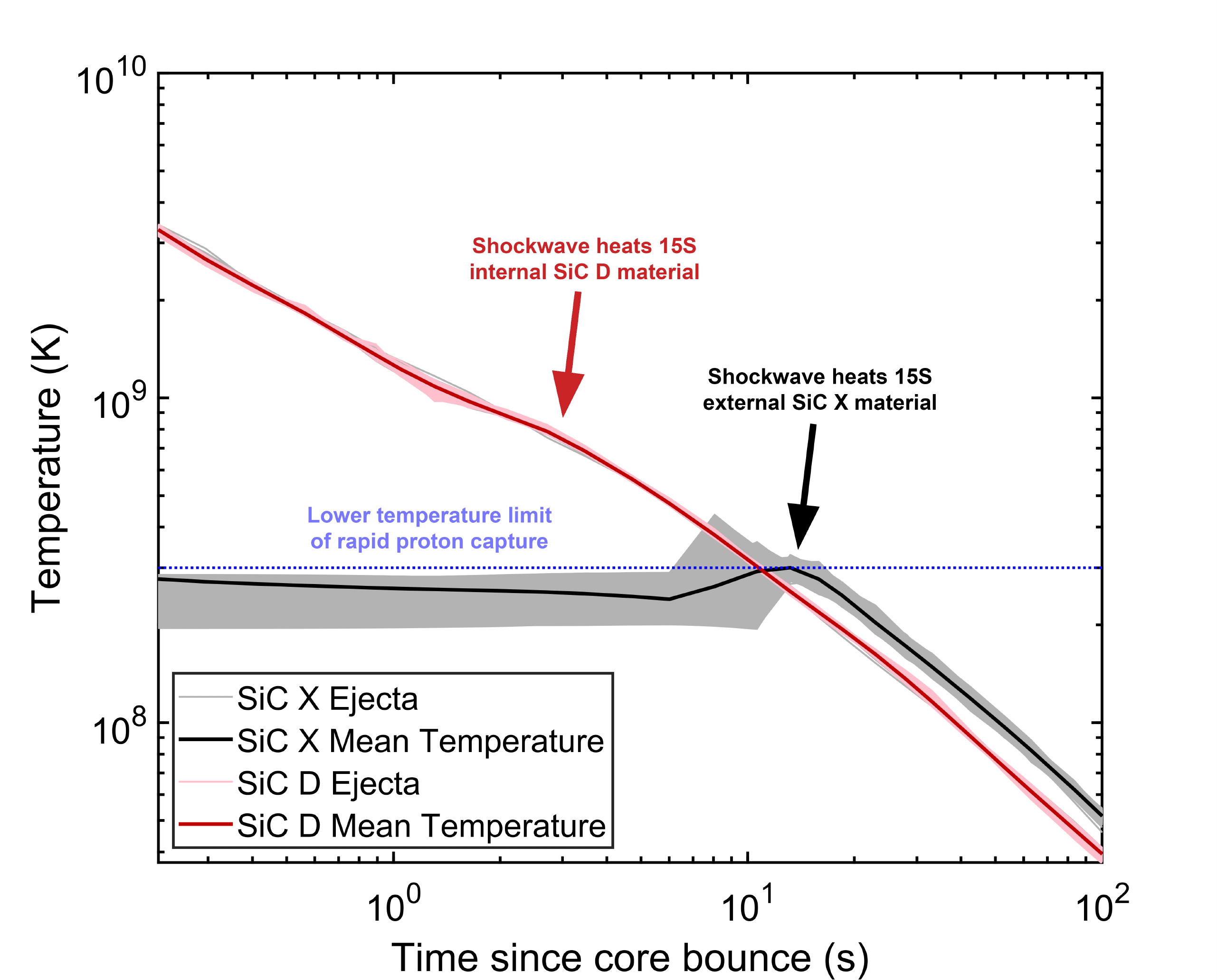}{0.55\textwidth}{(a) Temperature evolution of each model}
            \fig{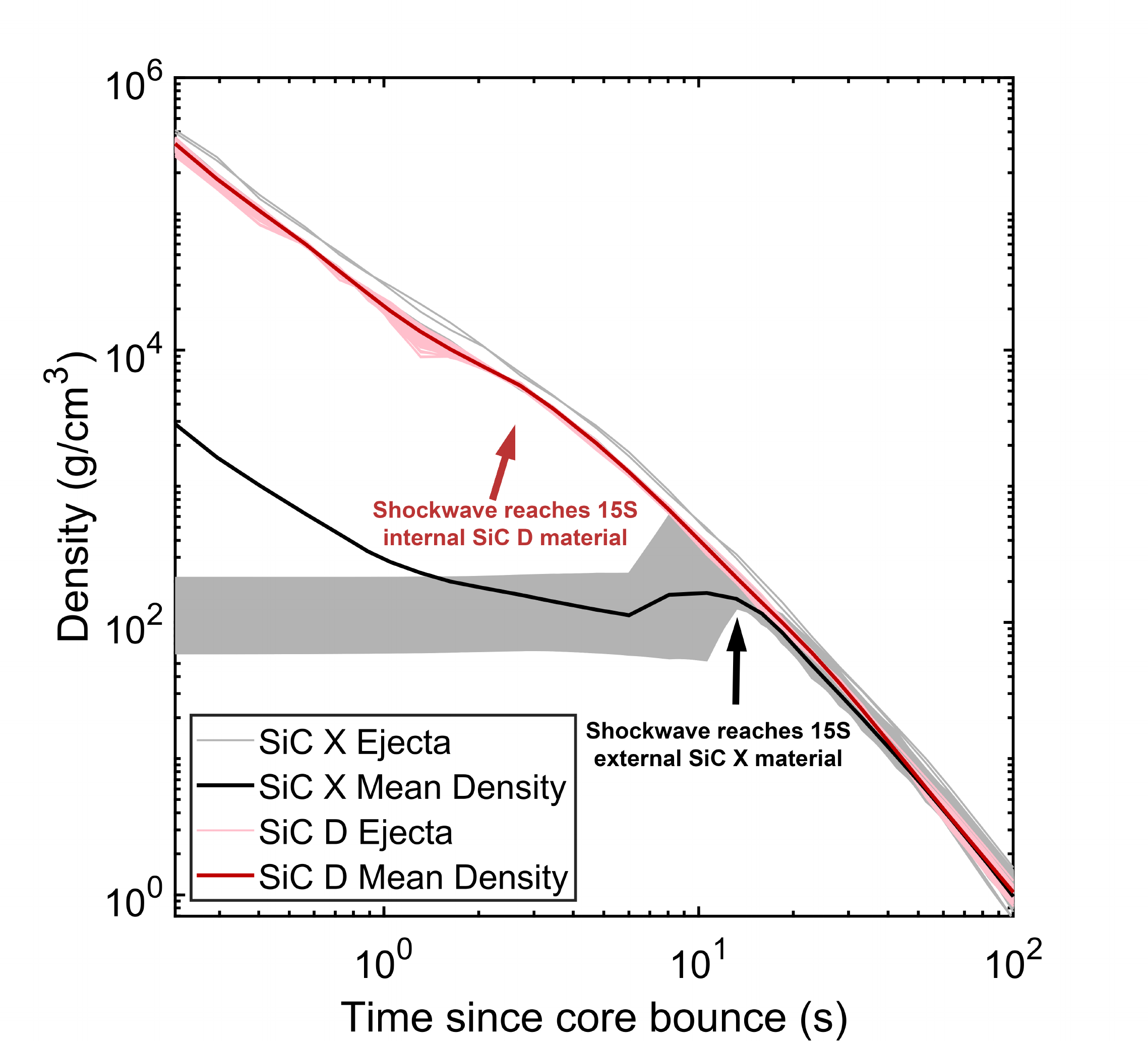}{0.5\textwidth}{(b) Density evolution of each model}}
  \caption{Shown are the temperature and density evolution of ejecta with carbon and nitrogen compositions similar to SiC D grains (red) and SiC X grains (black) in the first hundred seconds after core bounce. The minimum temperature above which rapid proton capture is expected to occur is presented as $3 \times 10^8$ K \citep{1998PhR...294..167S}. The supernova shockwave passes through the source material of SiC D grains first, and then reaches the SiC X grain source material. Average temperatures and densities were taken within limits defined by the $^{12}$C/$^{13}$C ratio of the source material. These limits were set by the upper and lower errors of the most extreme SiC D and X grains shown in Figure 2. SiC D grain source material has $^{12}$C/$^{13}$C $\in [1.4, 16.7]$ and SiC X grain source material has $^{12}$C/$^{13}$C $\in [10.2, 10046.5].$}
  \label{fig:TDHist}
\end{figure}

\begin{figure}[p]
  \centering
  \includegraphics[width=\linewidth]{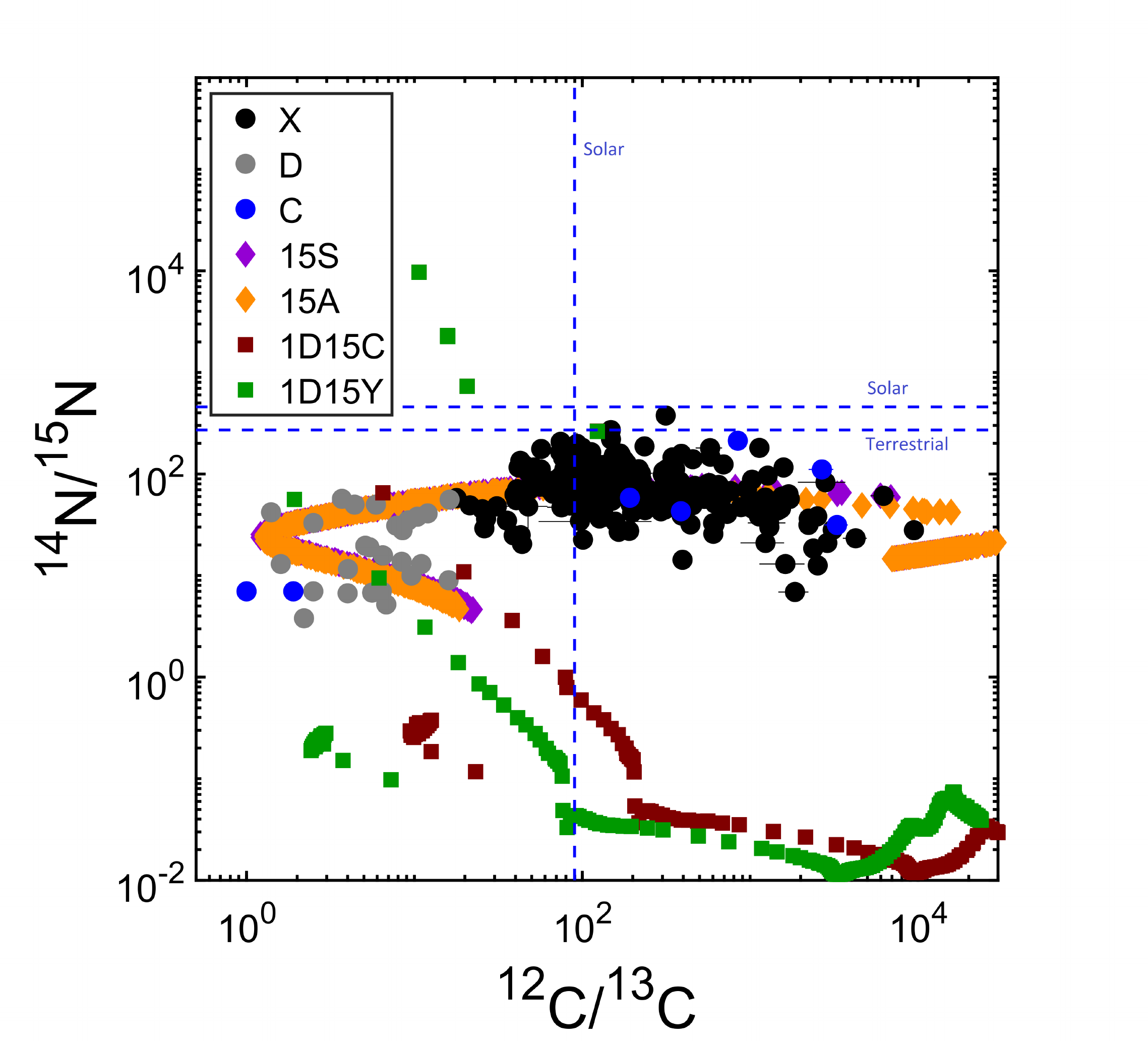}
  \caption{Carbon and nitrogen isotope yields from two post-explosion, 3-D 15 $M_\odot$ CCSN models and two 1-D 15 $M_\odot$ CCSN models from \cite{2002ApJ...576..323R} are plotted against SiC X, D, and C grain data retrieved from \citep{2009LPI....40.1198H,2020LPI....51.2140S,2019ApJ...873...14B,2008ApJ...689..622M}. Both 15S and 15A can explain the carbon and nitrogen signatures of all three grain types very accurately without any ad-hoc mixing between zones in a pre-supernova star for 1-D models.}
  \label{fig:CN}
\end{figure}

\begin{figure}
\gridline{\fig{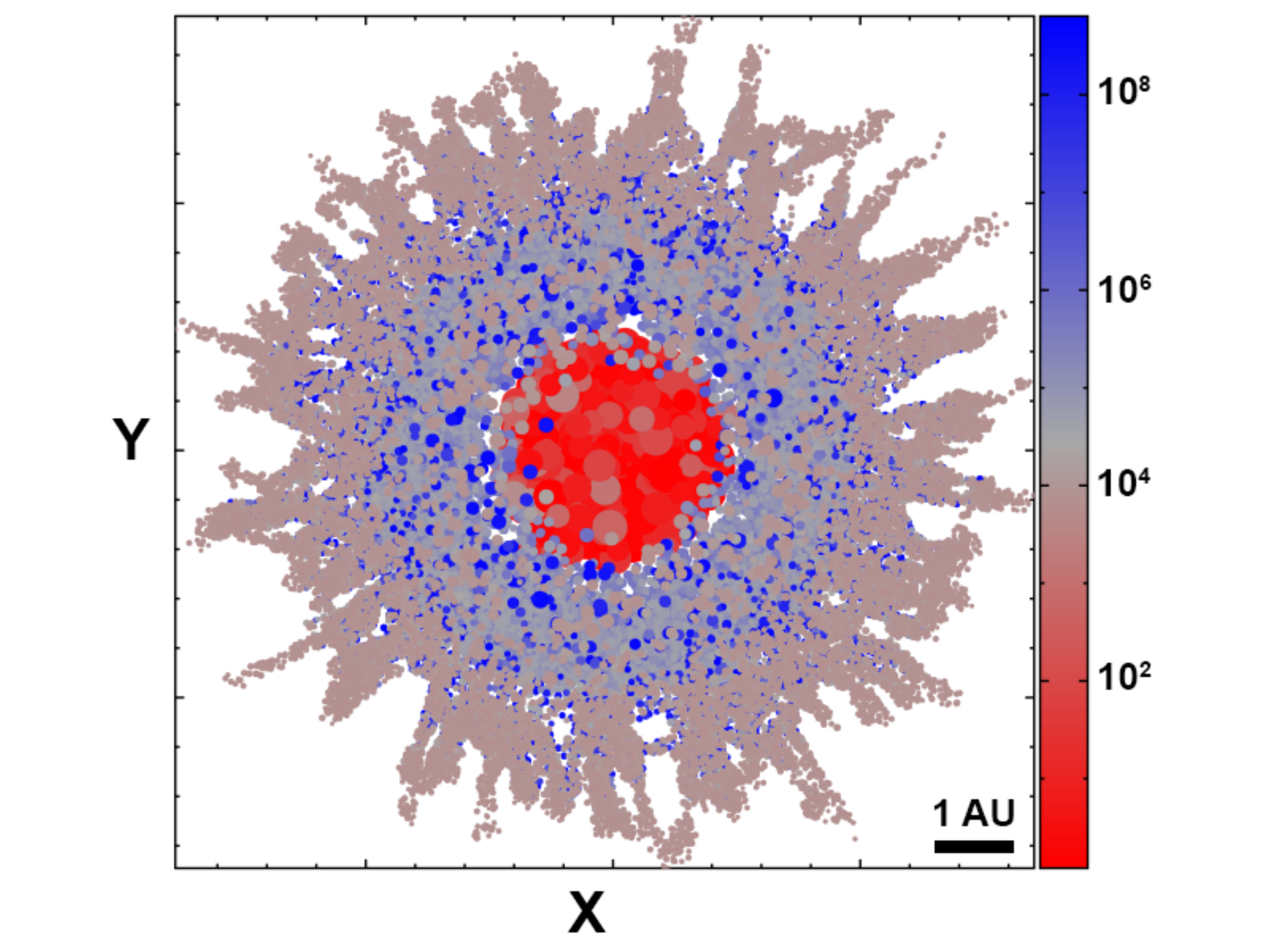}{0.24\textwidth}{(a) 15S $^{12}$C/$^{13}$C in X-Y plane}
          \fig{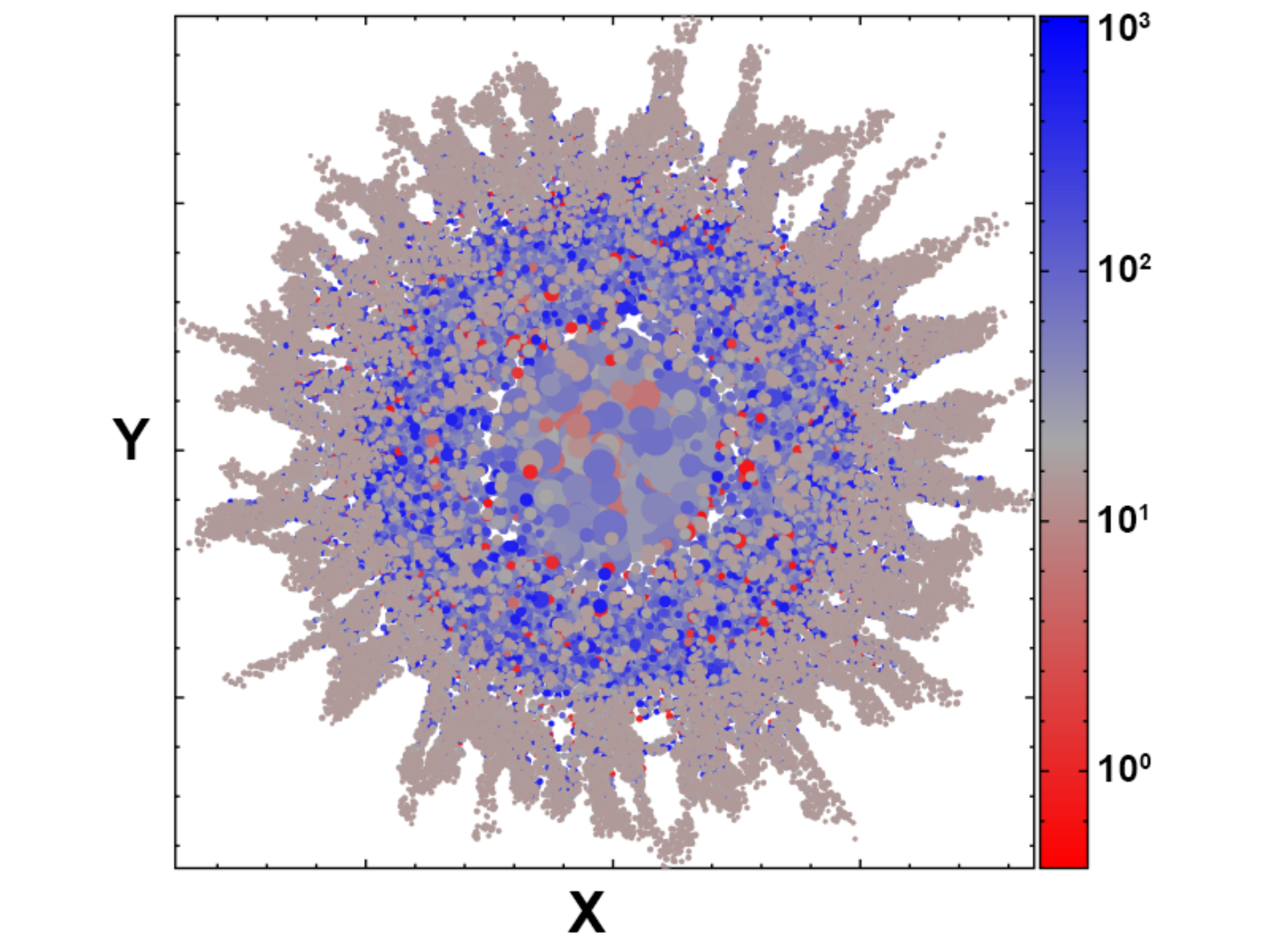}{0.24\textwidth}{(b) 15S $^{14}$N/$^{15}$N in X-Y plane}
          \fig{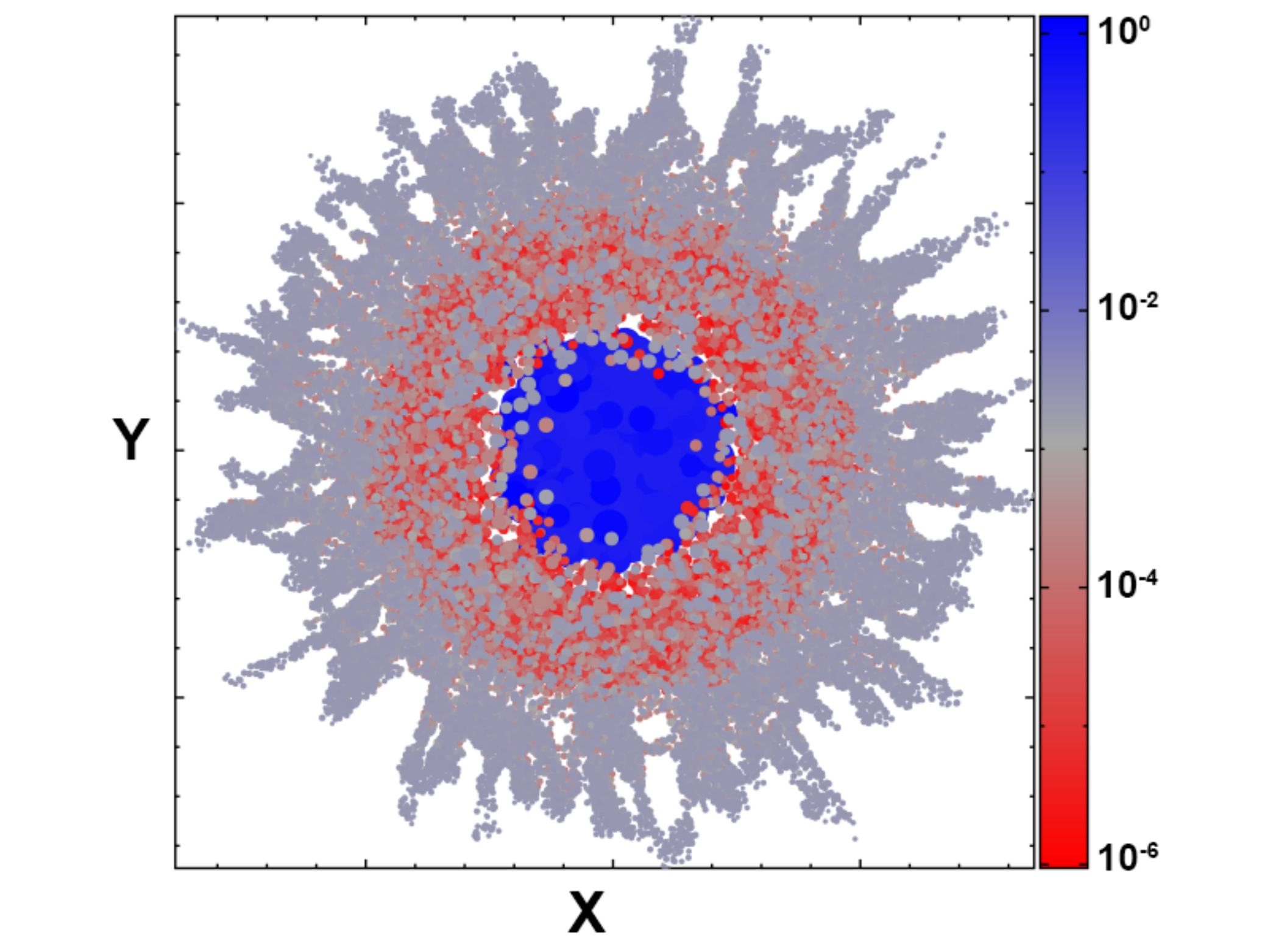}{0.24\textwidth}{(c) 15S $^{26}$Al/$^{27}$Al in X-Y plane}
          \fig{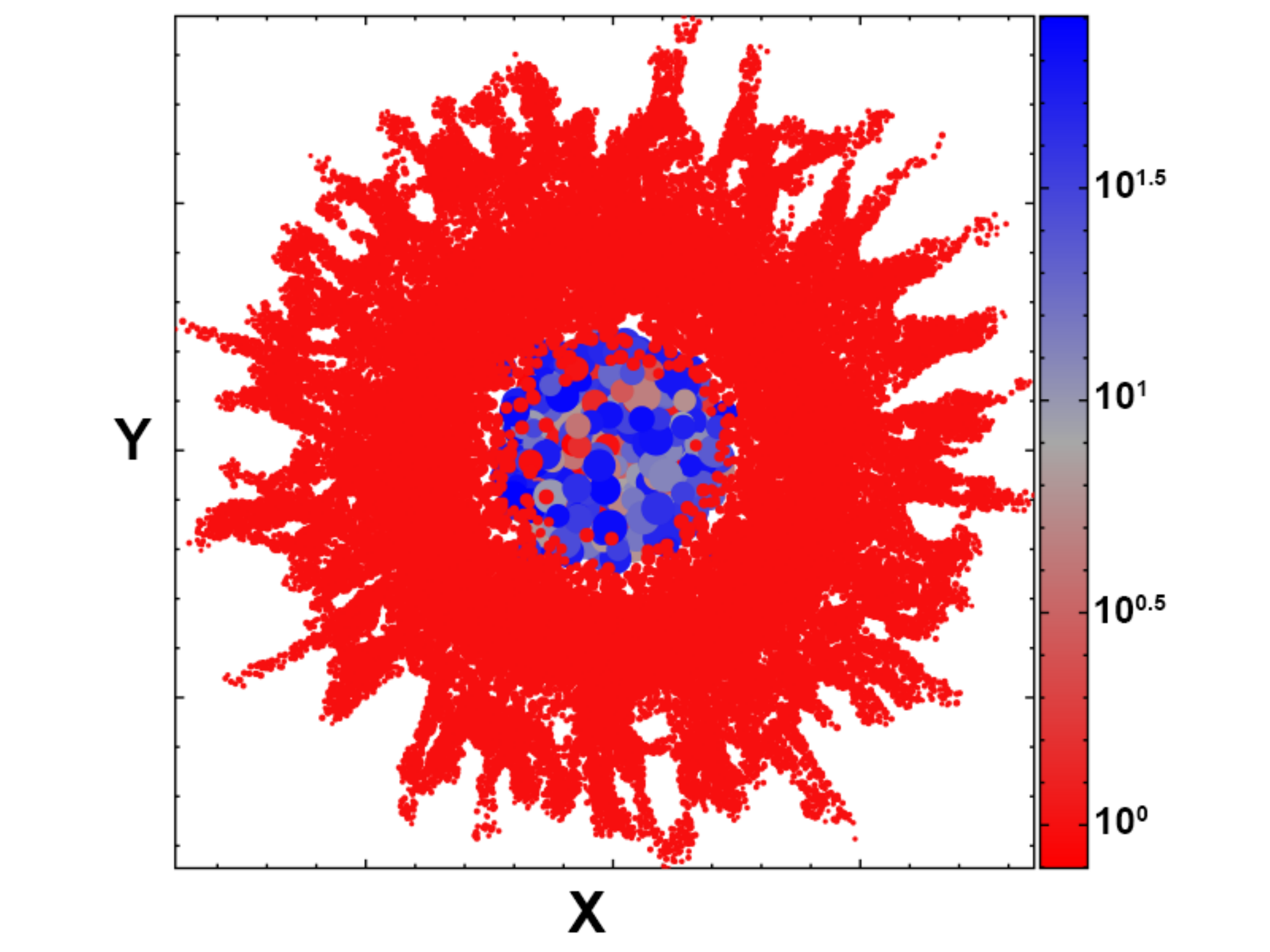}{0.24\textwidth}{(d) 15S $^{44}$Ti/$^{16}$O in X-Y plane}}
\gridline{\fig{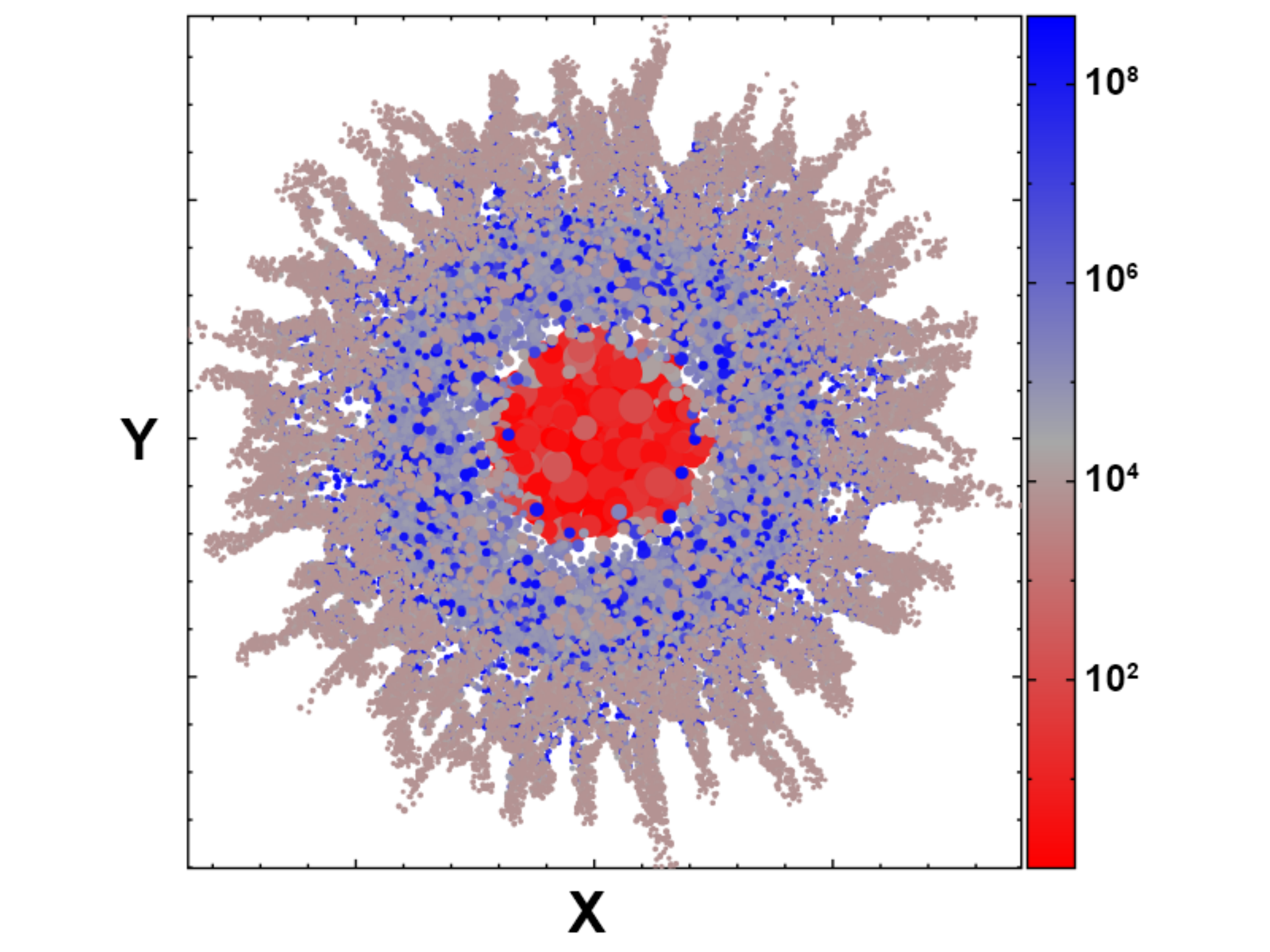}{0.24\textwidth}{(e) 15A $^{12}$C/$^{13}$C in X-Y plane}
          \fig{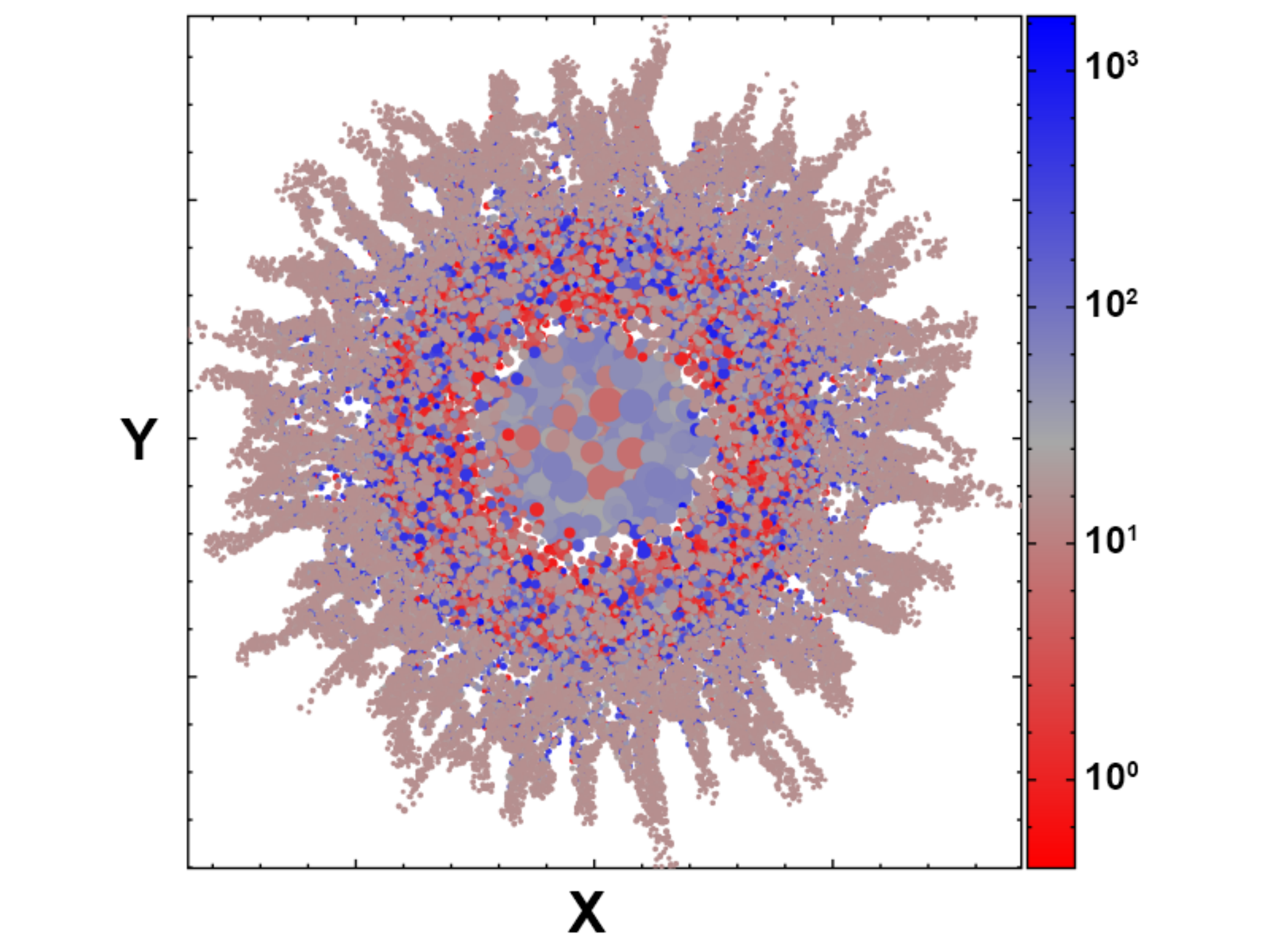}{0.24\textwidth}{(f) 15A $^{14}$N/$^{15}$N in X-Y plane}
          \fig{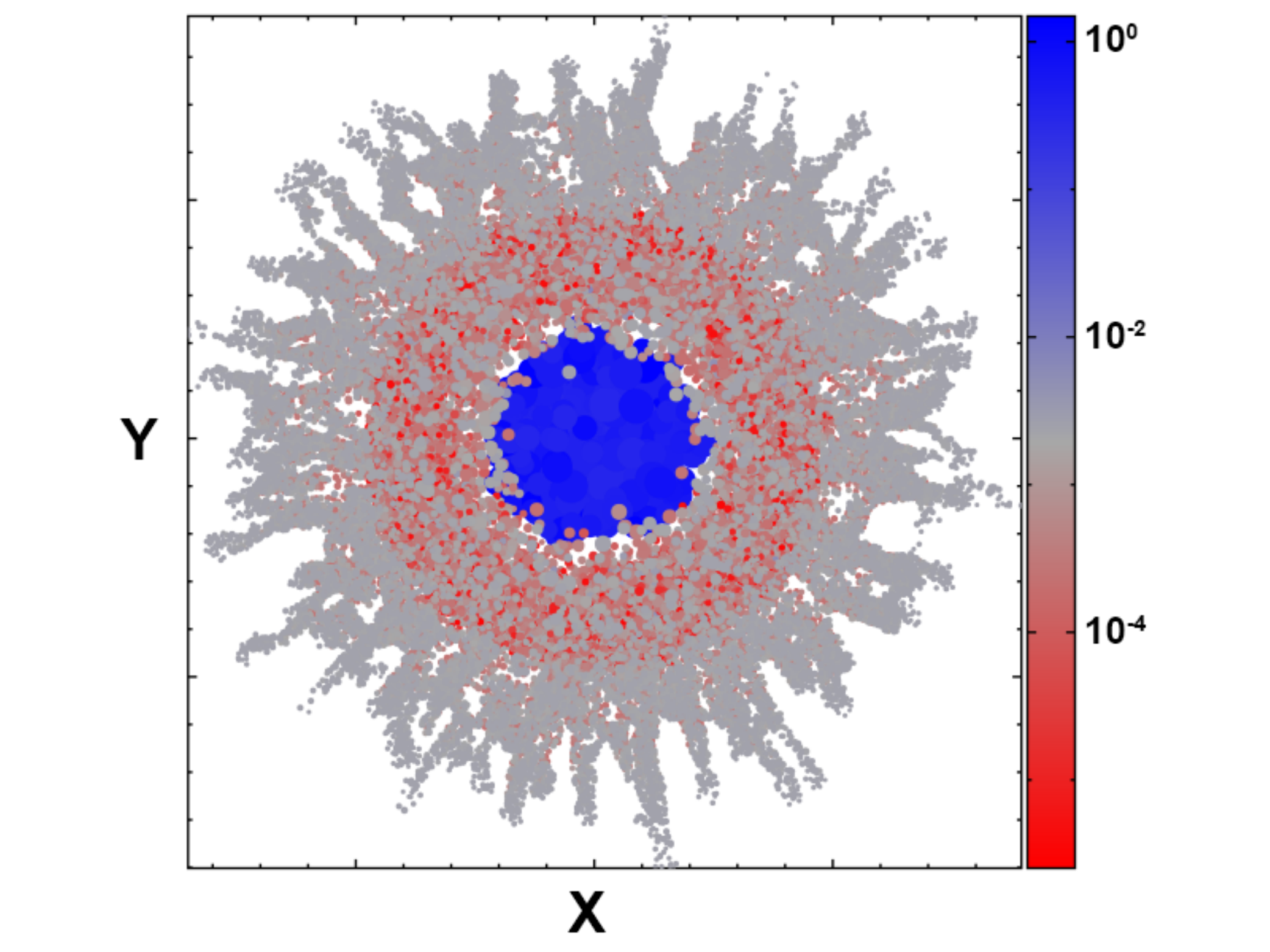}{0.24\textwidth}{(g) 15A $^{26}$Al/$^{27}$Al in X-Y plane}
          \fig{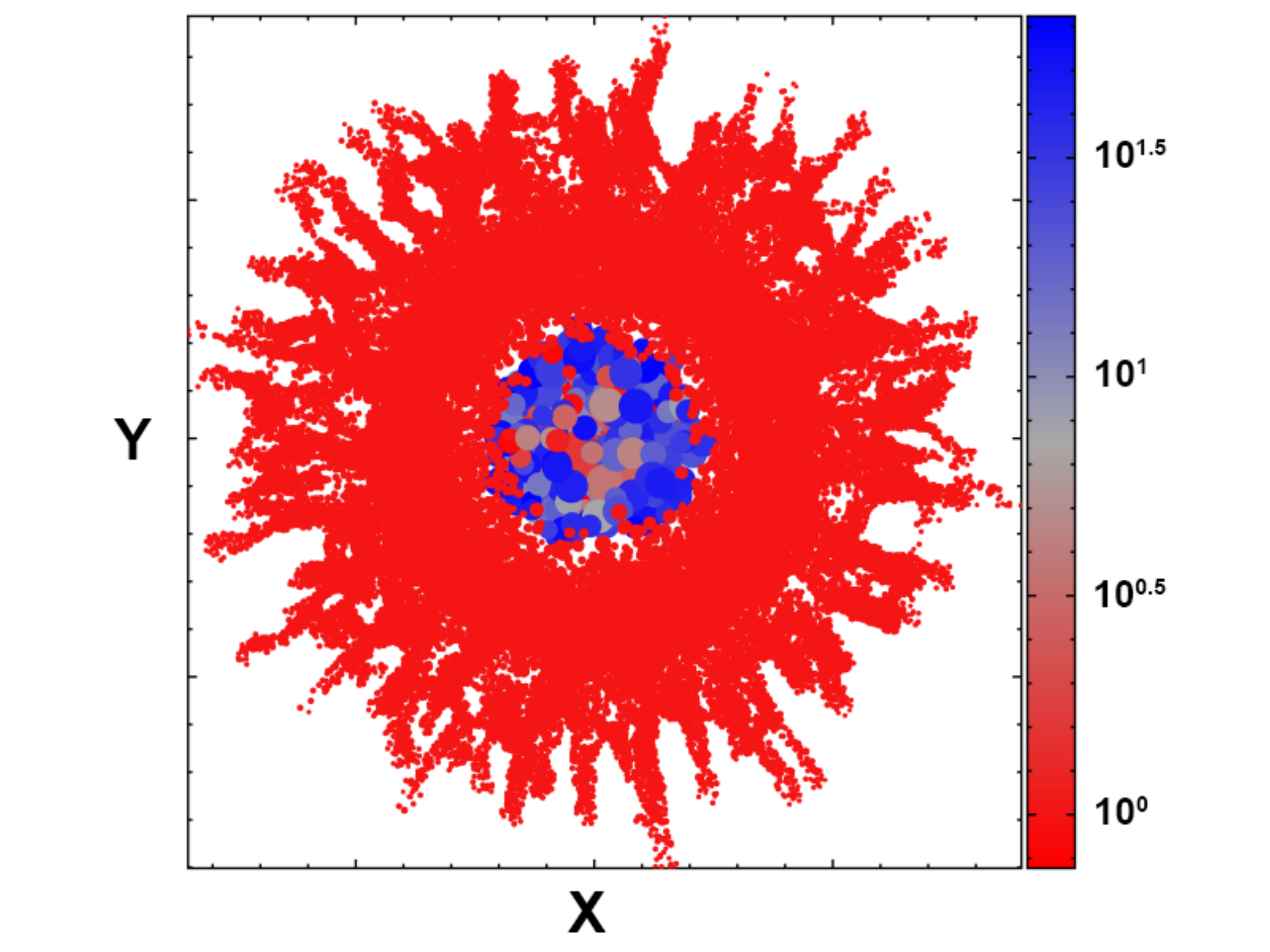}{0.24\textwidth}{(h) 15A $^{44}$Ti/$^{16}$O in X-Y plane}}
\gridline{\fig{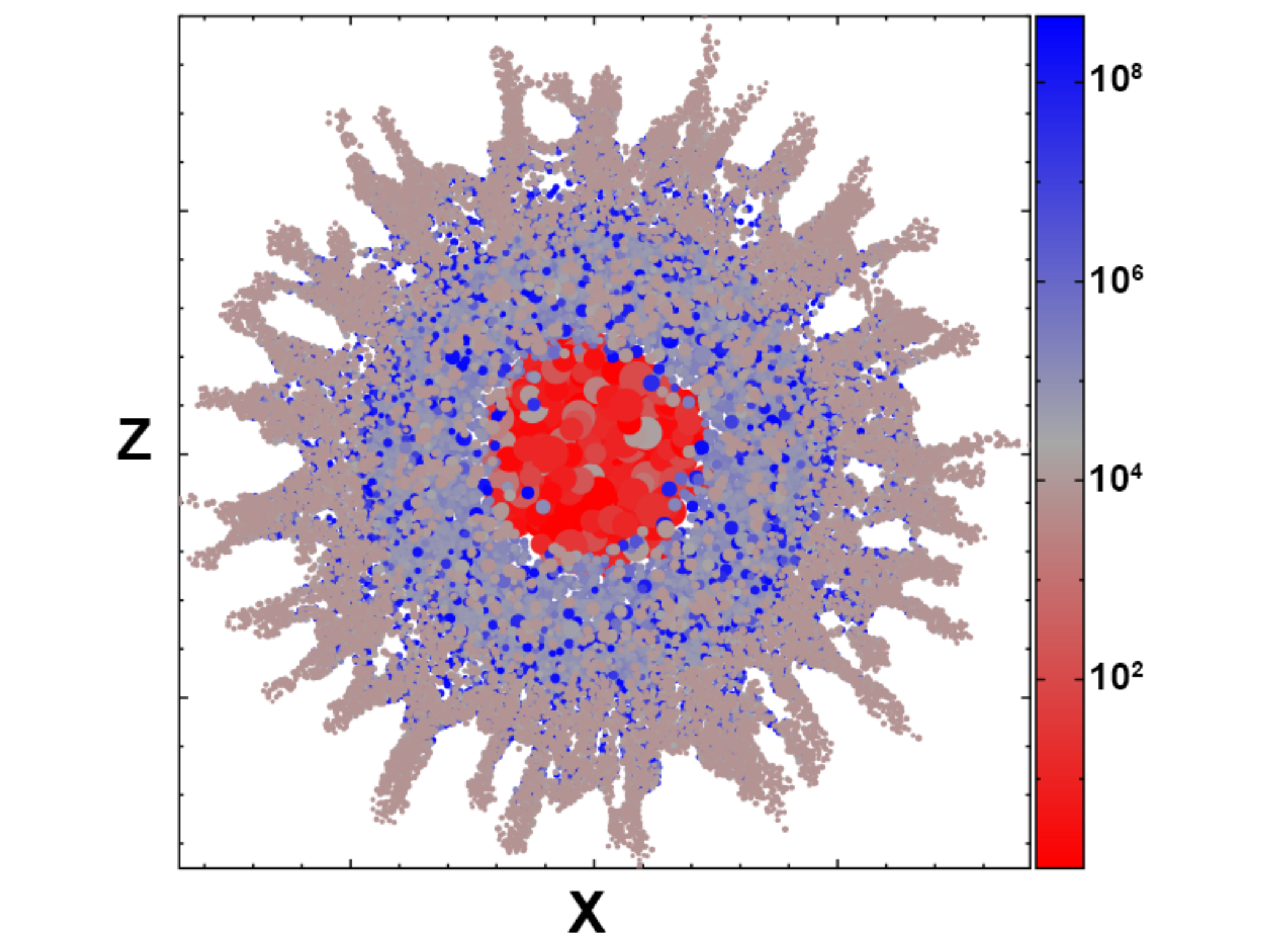}{0.24\textwidth}{(i) 15A $^{12}$C/$^{13}$C in X-Z plane}
          \fig{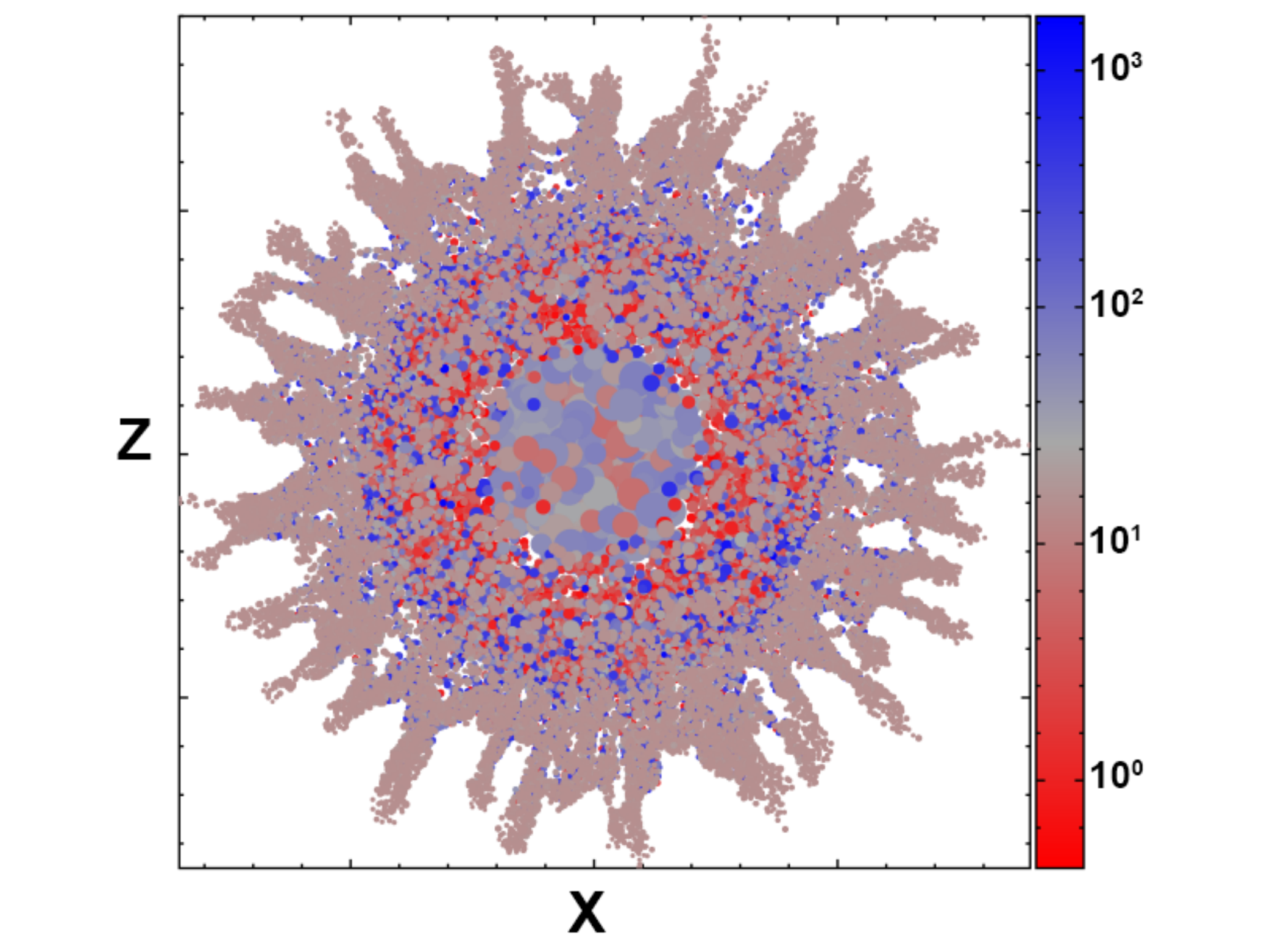}{0.24\textwidth}{(j) 15A $^{14}$N/$^{15}$N in X-Z plane}
          \fig{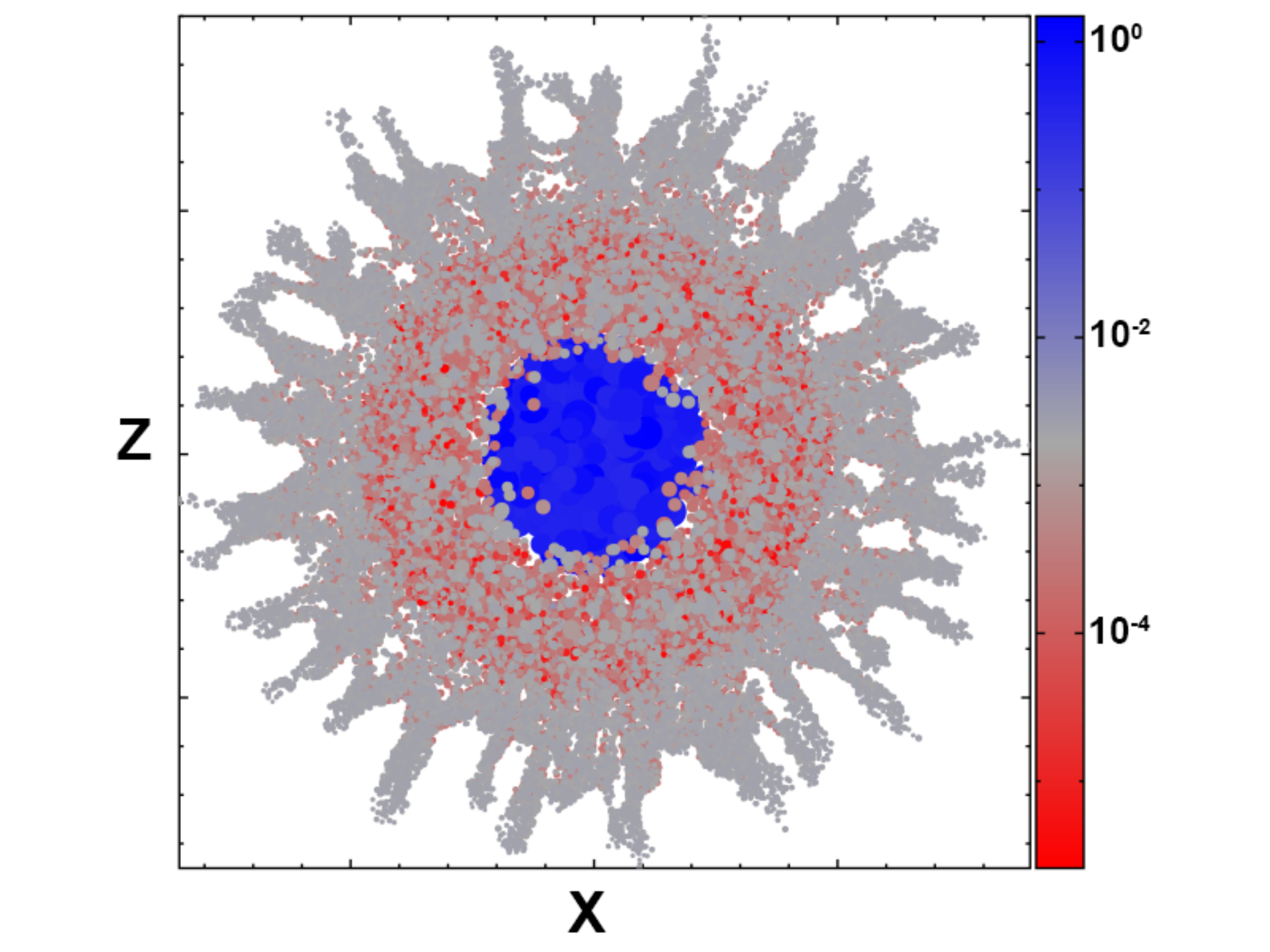}{0.24\textwidth}{(k) 15A $^{26}$Al/$^{27}$Al in X-Z plane}
          \fig{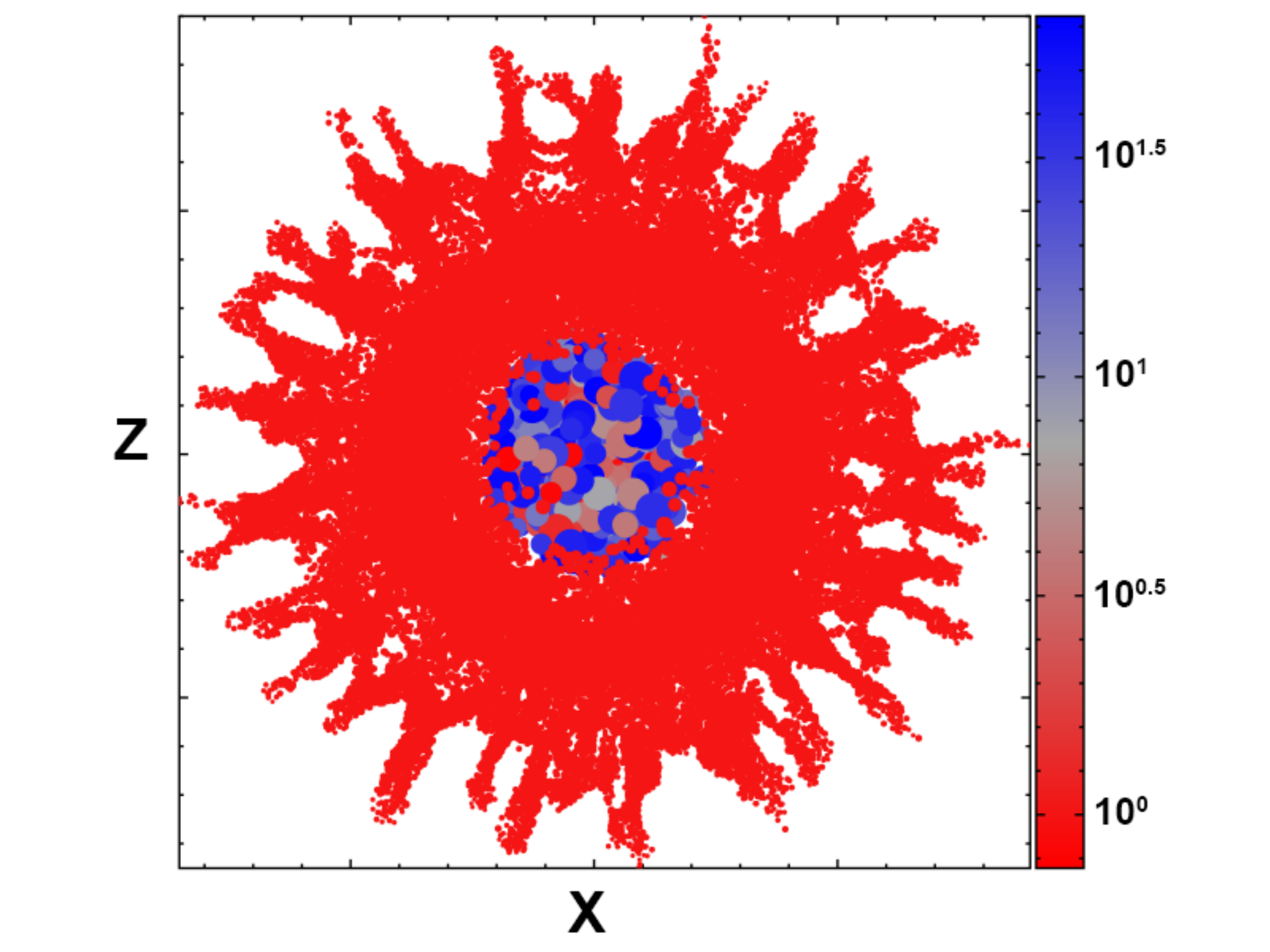}{0.24\textwidth}{(l) 15A $^{44}$Ti/$^{16}$O in X-Z plane}}
    \caption{Cross-sectional SPLASH plots of 3-D CCSN models 15S and 15A in the X-Y and X-Z planes showing the distribution of $^{12}$C/$^{13}$C, $^{14}$N/$^{15}$N, $^{26}$Al/$^{27}$Al, and $^{44}$Ti/$^{16}$O isotope ratios, 43 hours after core-collapse. All plots represent the same spatial scales, shown here in astronomical units (AU). Marker sizes are scaled by $0.7 \times$ the smoothing length of each clump. The colorbar represents the logarithm of the ratios of each set of two isotopes as labeled above, where blue shades correspond to higher ratios and red shades correspond to lower ratios. The innermost material of both CCSN explosions is rich in $^{13}$C, $^{15}$N, $^{26}$Al, and $^{44}$Ti. All of these signatures are consistent with the isotope signatures of SiC D grains and imply that SiC D grains likely condense from material deep within the interior of a 15 $M_\odot$ CCSN.}
    \label{fig:SPLASH}
\end{figure}

\begin{figure}[p]
\centering
\includegraphics[width = \linewidth]{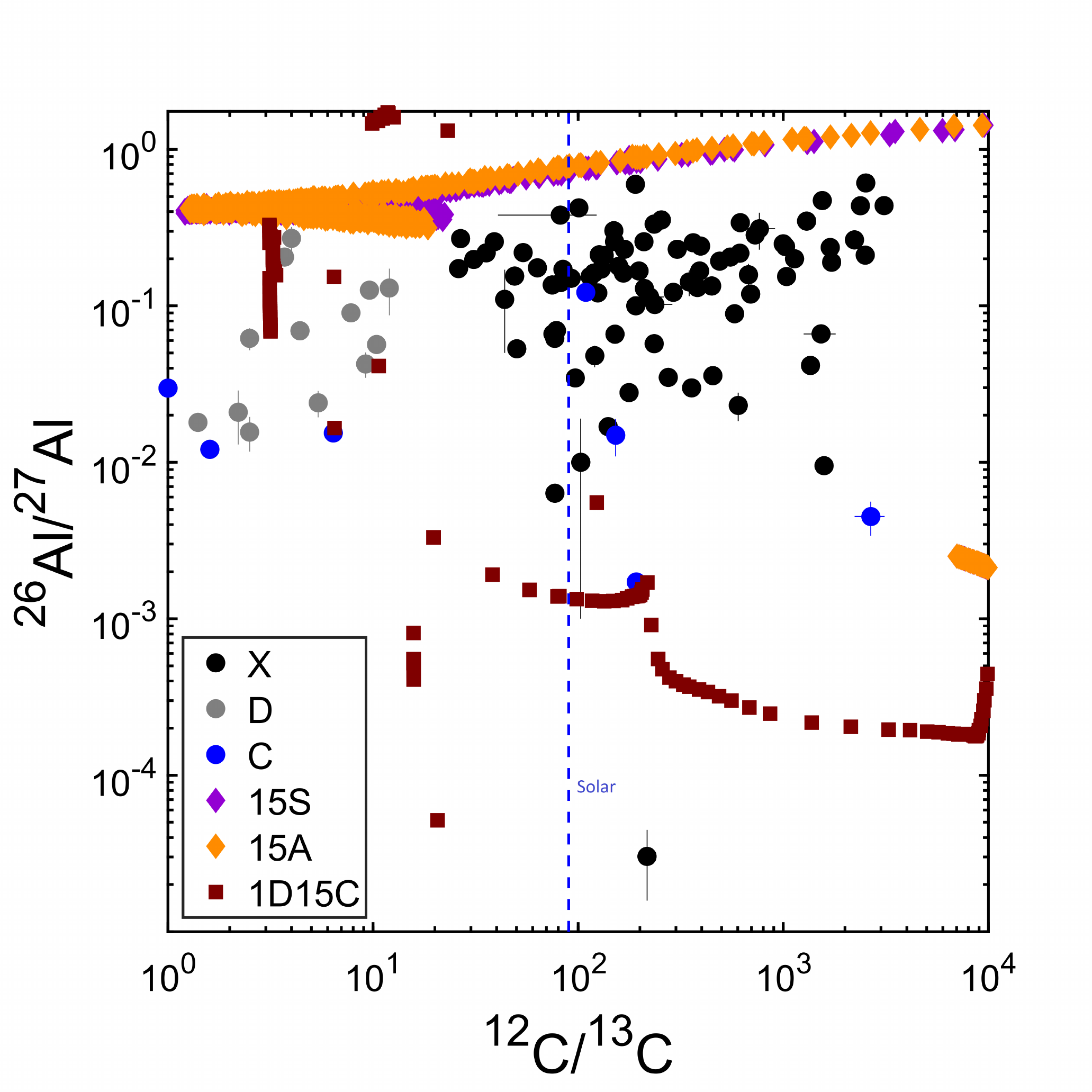}
\caption{Aluminum compositions of nucleosynthetic models 15S, 15A, and 1D15C compared to implied aluminum ratios in SiC X, D, and C grains \citep{1992ApJ...394L..43A,1996ApJ...462L..31N,1997GeCoA..61.5117H,2001ApJ...551.1065A,2002ApJ...575..257L,2005ApJ...631L..89N,2007GeCoA..71.4786Z,2010ApJ...719.1370H,2012ApJ...745L..26H,2015ApJ...799..156X,2018GeCoA.221...60G}. Both 3-D models more accurately predict the $^{26}$Al excesses of SiC X, D, and C grains than the 1-D model. The slightly larger $^{26}$Al/$^{27}$Al ratios in 3-D CCSN model yields compared to stardust may be remedied by considering minor contamination from $^{26}$Al-poor material in the stellar envelope.}
\label{fig:Al}
\end{figure}

\begin{figure}[p]
  \centering
  \includegraphics[width=\linewidth]{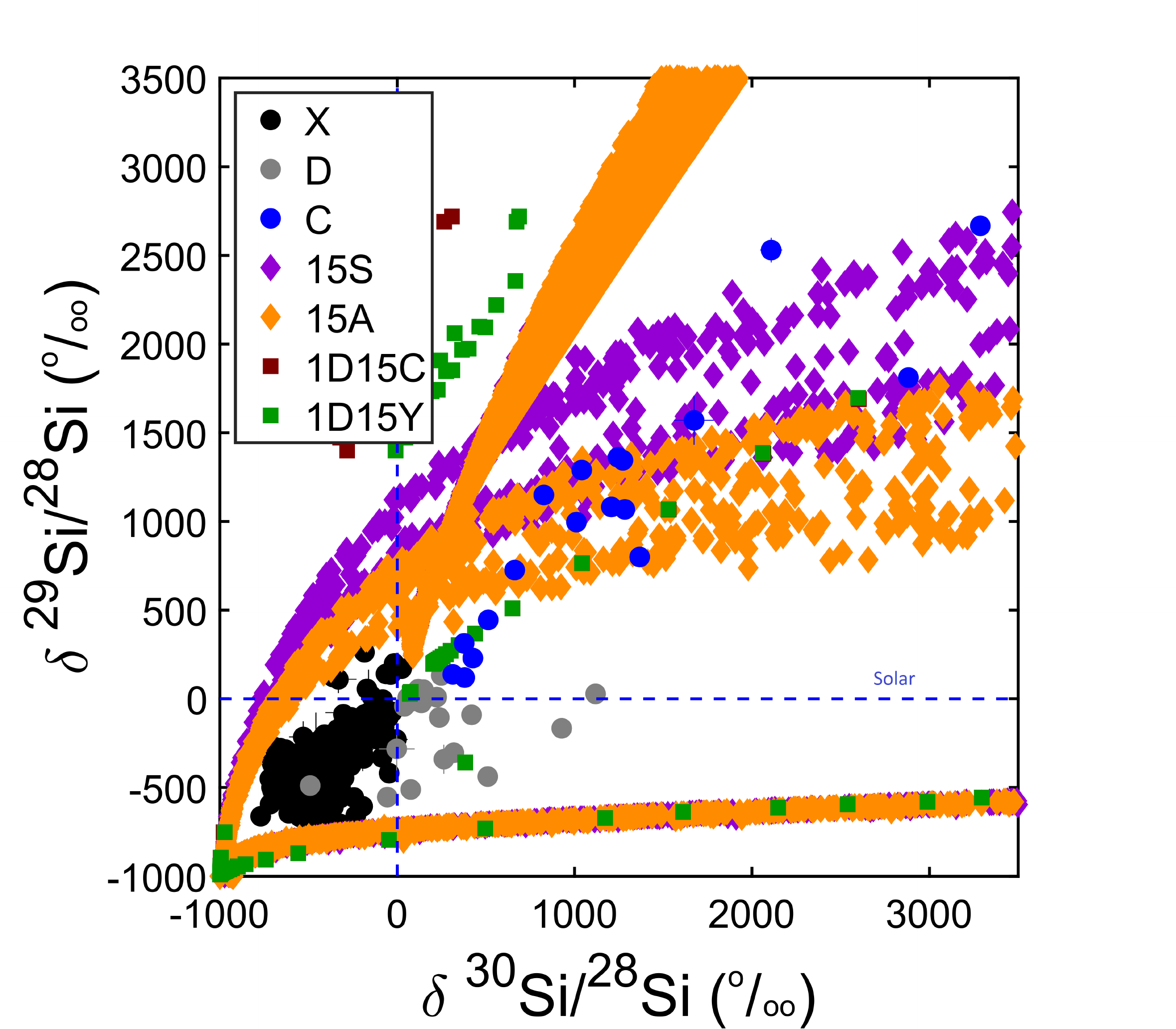}
  \caption{Silicon isotope yields from two post-explosion, 3-D 15 $M_\odot$ CCSN models and two 1-D 15 $M_\odot$ CCSN models plotted against SiC X, D, and C grain data retrieved from \citep{2009LPI....40.1198H,2020LPI....51.2140S,2019ApJ...873...14B,2008ApJ...689..622M}. The 3-D models predict the Si compositions of SiC C grains without considering any mixture but necessitate mixture between ejecta and SN envelope to explain SiC X and possibly D grains. Clumps with $^{28}$Si excesses are widespread in the simulations.}
  \label{fig:Si}
\end{figure}

\begin{figure}
\centering
\includegraphics[width=\linewidth]{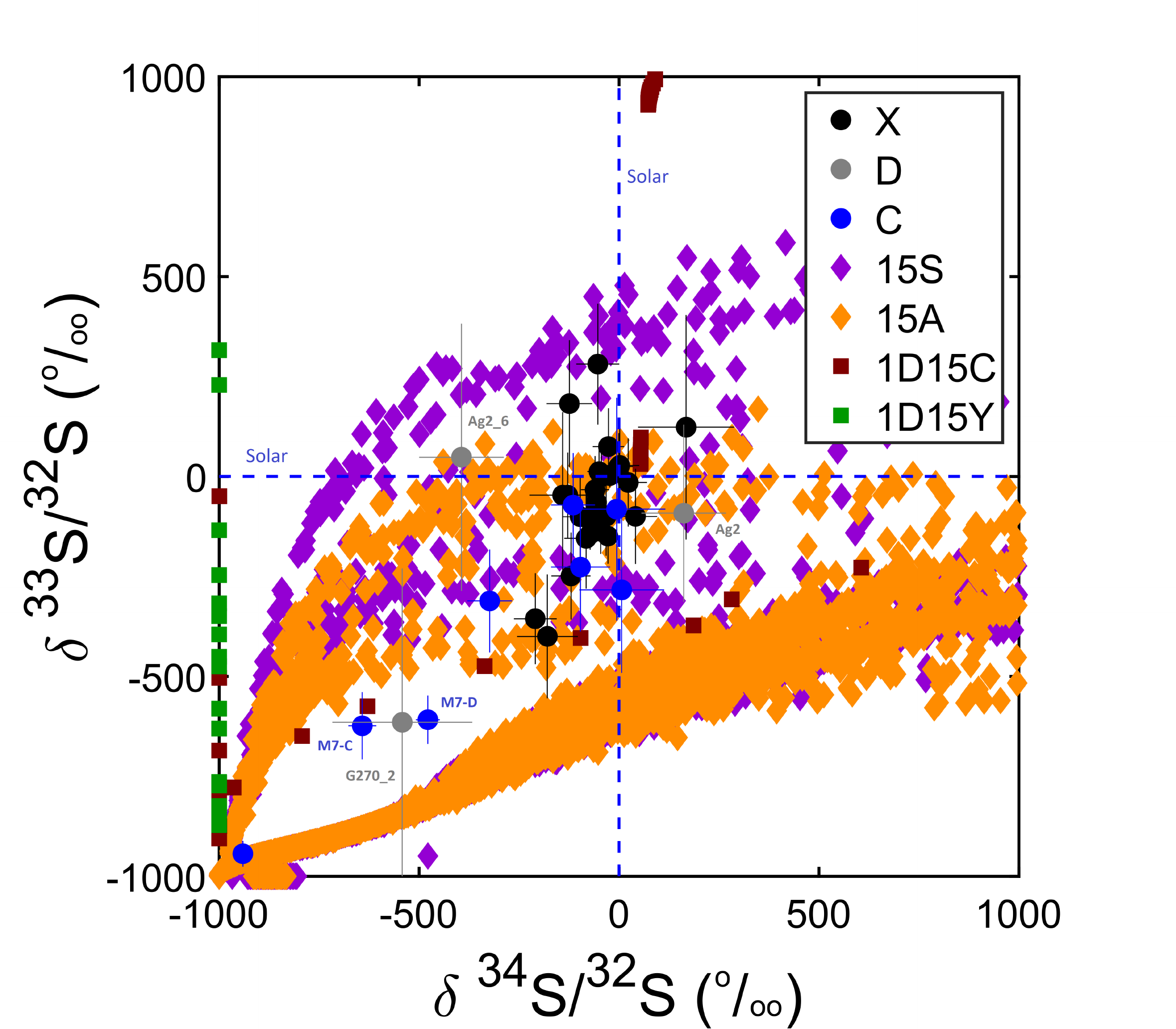}
\caption{Sulfur isotope yields of 3-D CCSN models 15S and 15A 43 hours after core-collapse compared to SiC stardust. SiC stardust isotopic data were collected from \citep{2010ApJ...717..107G,2012ApJ...745L..26H,2012LPI....43.2679O,2015ApJ...799..156X,2016ApJ...820..140L}. SiC C grains from \cite{2012ApJ...745L..26H} which do not fit the predictions of 15S and 15A, along with SiC D grains from \cite{2016ApJ...820..140L}, are labeled using the identifiers used in their respective works. Clumps with enrichments of $^{32}$S, which explain the majority of SiC stardust, are found to be farther than 1.4 AU away from the center of the explosion, while clumps with depletions of $^{32}$S (and enrichments of the neutron-rich isotopes $^{33}$S and $^{34}$S) are less than 1.4 AU from the center of the explosion.}
\label{fig:S}
\end{figure}

\begin{figure}
\begin{rotatetable}
\gridline{\fig{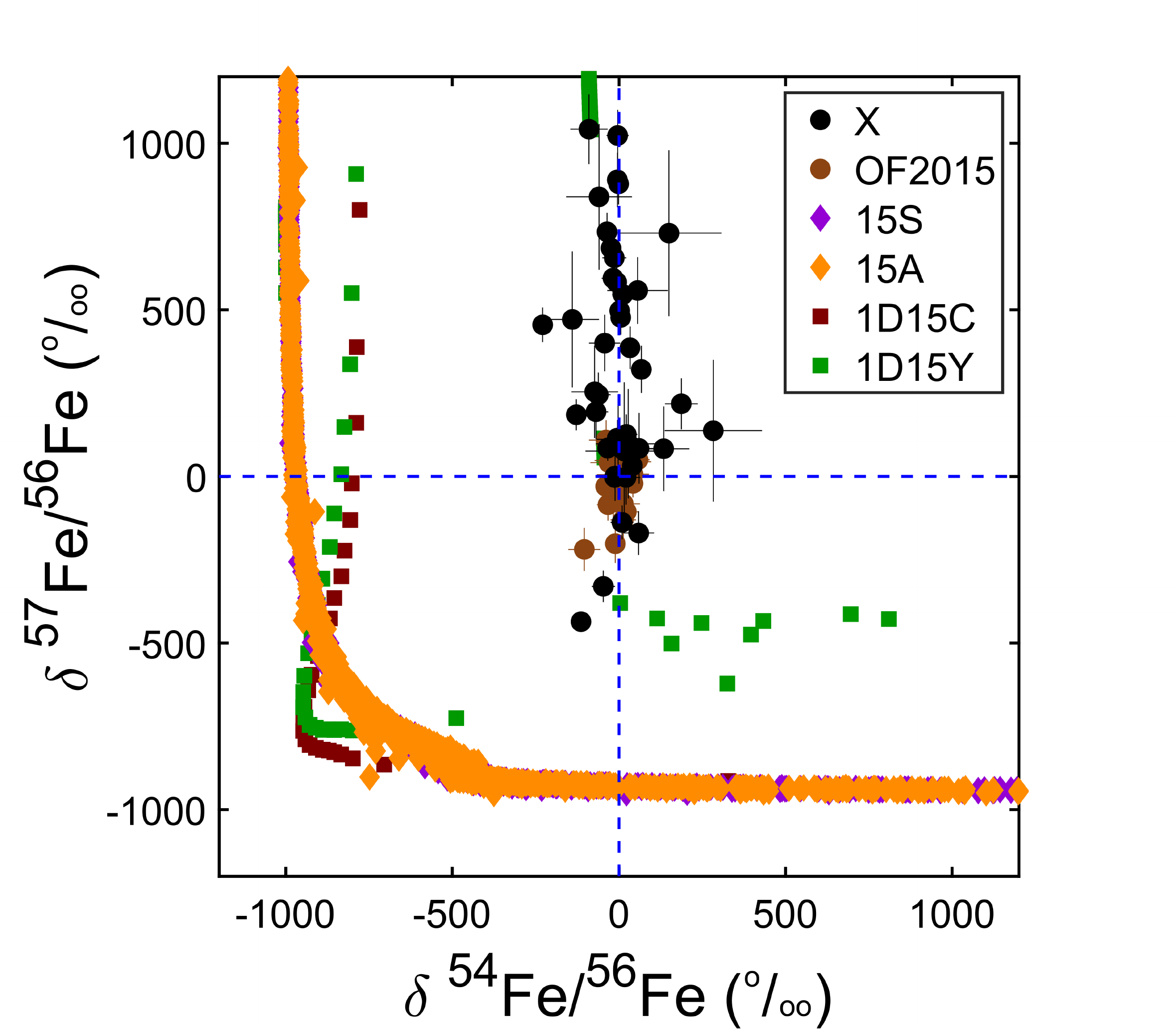}{0.45\textwidth}{(a) Fe Isotopic Comparison to SiC stardust}}
\gridline{\fig{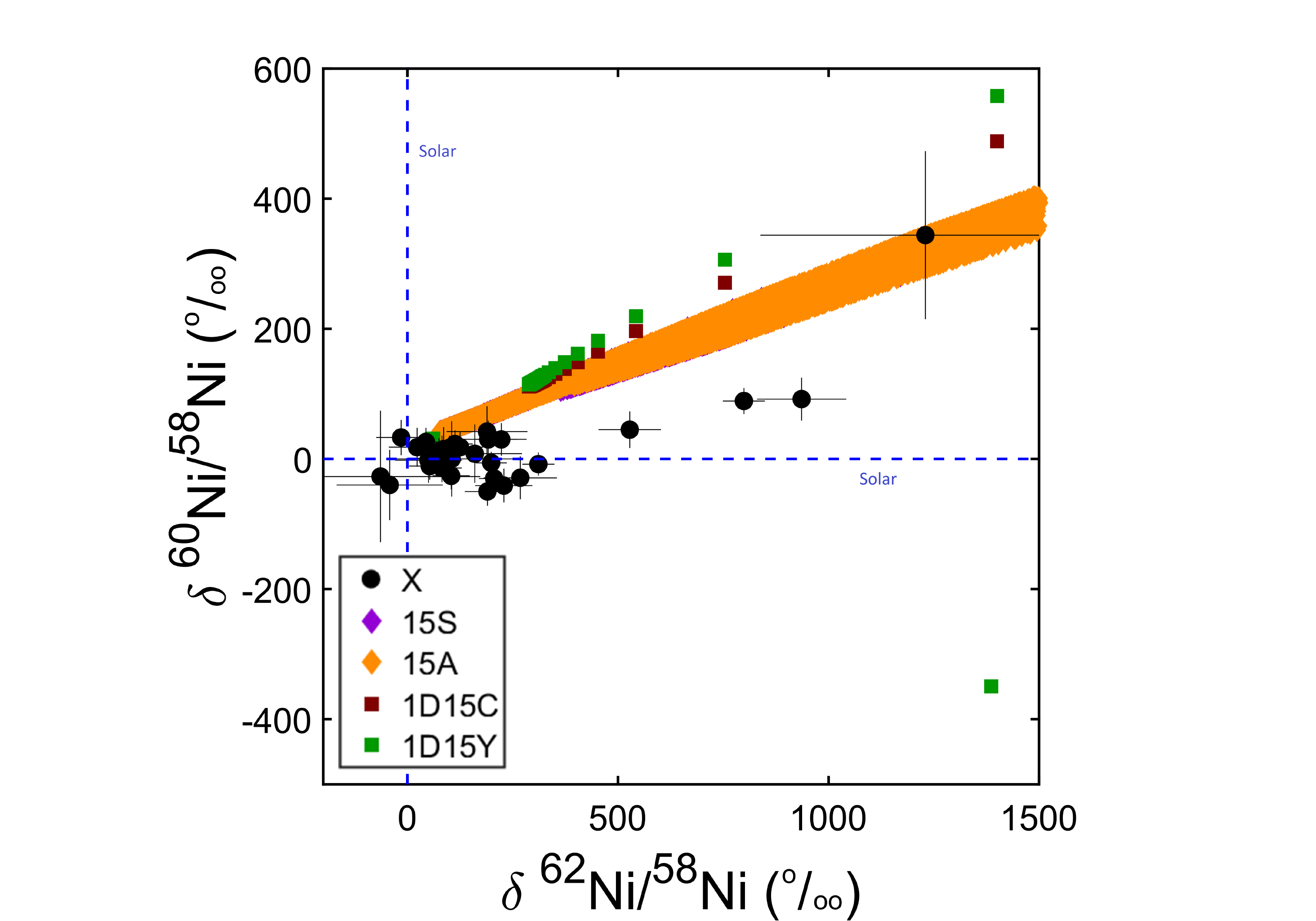}{0.55\textwidth}{(b) Ni Isotopic Comparison to SiC stardust}
          \fig{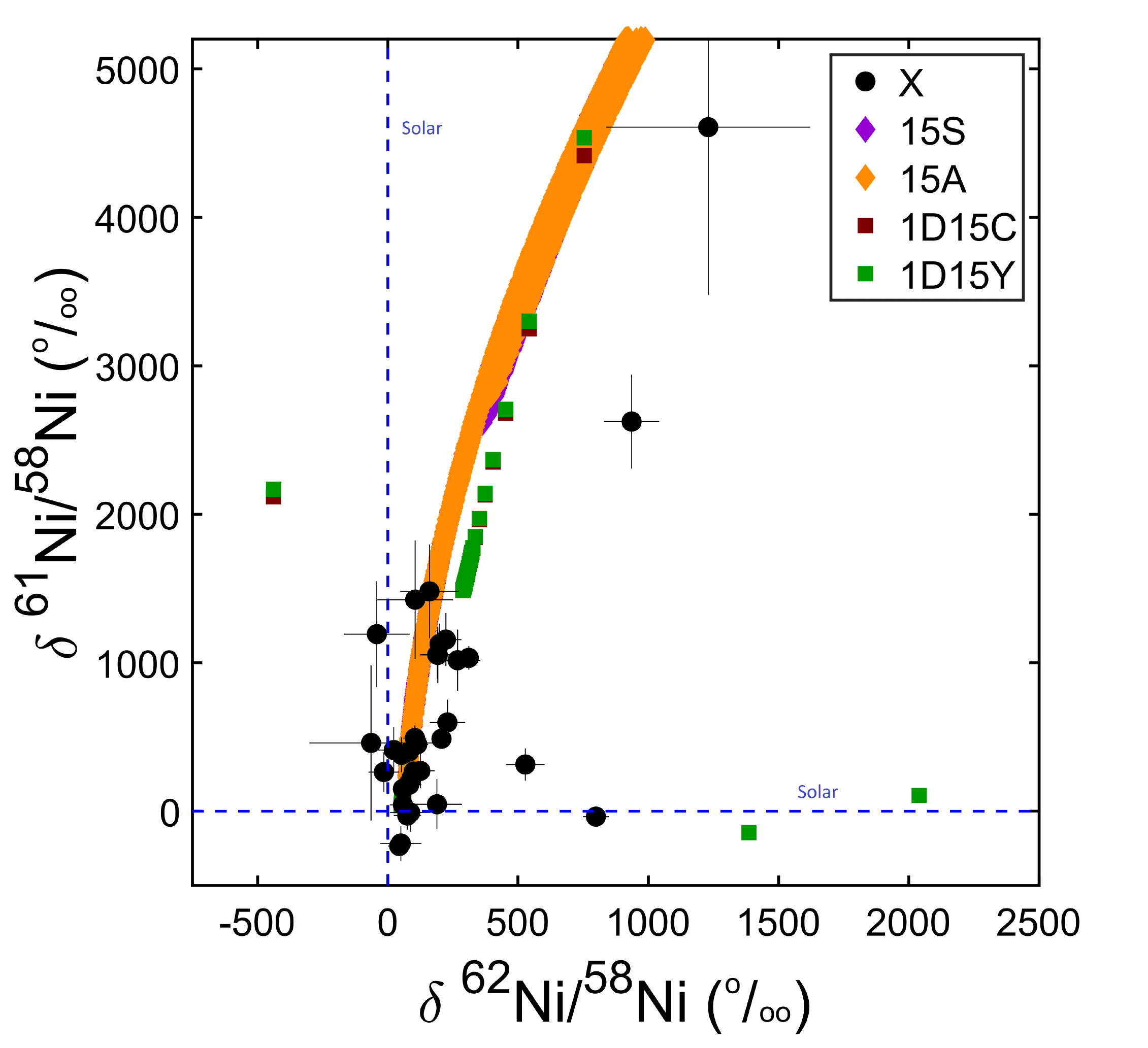}{0.4\textwidth}{(c) Ni Isotopic Comparison to SiC stardust}}
    \caption{Iron and Nickel isotopic yields compared to SiC stardust. SiC stardust isotopic data were collected from \cite{2008ApJ...689..622M}. "OF2015" in (a) refers to oxide and silicate stardust from \cite{2015M&PS...50.1392O}.} 3-D simulations 15S and 15A, which show almost identical Fe and Ni signatures, work slightly better to explain SiC X grain compositions in the case of each isotope of Fe and Ni. Large enrichments of $^{57}$Fe can be found deep within the interior of either 3-D CCSN model.
    \label{fig:FeNi}
\end{rotatetable}
\end{figure}

\end{document}